\documentclass[preprint,journal]{vgtc}
\ifpdf                                
  \pdfoutput=1\relax                   
  \usepackage{graphicx}               
  \DeclareGraphicsExtensions{.pdf,.png,.jpg,.jpeg} 
\else
  \usepackage{graphicx}                
  \DeclareGraphicsExtensions{.eps}     
\fi
\usepackage{microtype}                 
\PassOptionsToPackage{warn}{textcomp}  
\usepackage{textcomp}                  
\usepackage{mathptmx}                  
\usepackage{times}                    
         
\usepackage{cite}                      
\usepackage{tabu}                      
\usepackage{booktabs}                  

\graphicspath{{./figs/}}

\usepackage[dvipsnames]{xcolor}
\usepackage{color}

\usepackage{graphicx}

\usepackage{amssymb}
\usepackage{amsmath}
\usepackage{amsthm} 
\usepackage{amsfonts}

\usepackage{bm}

\newcommand {\mm}[1] {\ifmmode{#1}\else{\mbox{\(#1\)}}\fi}

\newcommand{\Rspace}        {\mm{\mathbb{R}}}
\newcommand{\Xspace}        {\mm{\mathbb{X}}}

\newcommand{\IM}        {\mm{\mathcal{D}}}

\newcommand{\OP}        {\mm{\rm{\bf OP\_6}}}

\newcommand{\T}        {\mm{\mathcal{T}}}

\newcommand{\LCA}        {\mm{a}}

\newcommand{\MT}{\mm{\rm MergeTree\,}}

\newcommand{\myemph}[1]{\textbf{#1}}

\newcommand{\sm}  {\mbox{sim}}

\newtheorem{theorem}{Theorem}[section]
\theoremstyle{definition}
\newtheorem{definition}{Definition}[section]

\newcommand{\denselist}{\vspace{-5pt} \itemsep -2pt\parsep=-1pt\partopsep -2pt}

\newcommand{\para}[1]        {\vspace{2mm}\noindent{\textbf{#1}}}

\newcommand*{\img}[1]{%
    \raisebox{-.3\baselineskip}{%
        \includegraphics[
        height=\baselineskip,
        width=\baselineskip,
        keepaspectratio,
        ]{#1}%
    }%
}

\ieeedoi{10.1109/TVCG.2019.2934242}

\onlineid{1041}

\vgtccategory{Research}

\vgtcinsertpkg

\usepackage{todonotes}

\title{A Structural Average of Labeled Merge Trees \\for Uncertainty Visualization}

\author{Lin Yan, Yusu Wang, Elizabeth Munch, Ellen Gasparovic, and Bei Wang}
\authorfooter{
\item
Lin Yan is with University of Utah. E-mail: lin.yan@utah.edu.
\item
Yusu Wang is with Ohio State University. E-mail: yusu@cse.ohio-state.edu. 
\item
Elizabeth Munch is with Michigan State University. E-mail: muncheli@egr.msu.edu.
\item
Ellen Gasparovic is with Union College. E-mail: gasparoe@union.edu.
\item
Bei Wang is with University of Utah. E-mail: beiwang@sci.utah.edu.
}

\teaser{
 \centering
  \includegraphics[width=1.0\columnwidth]{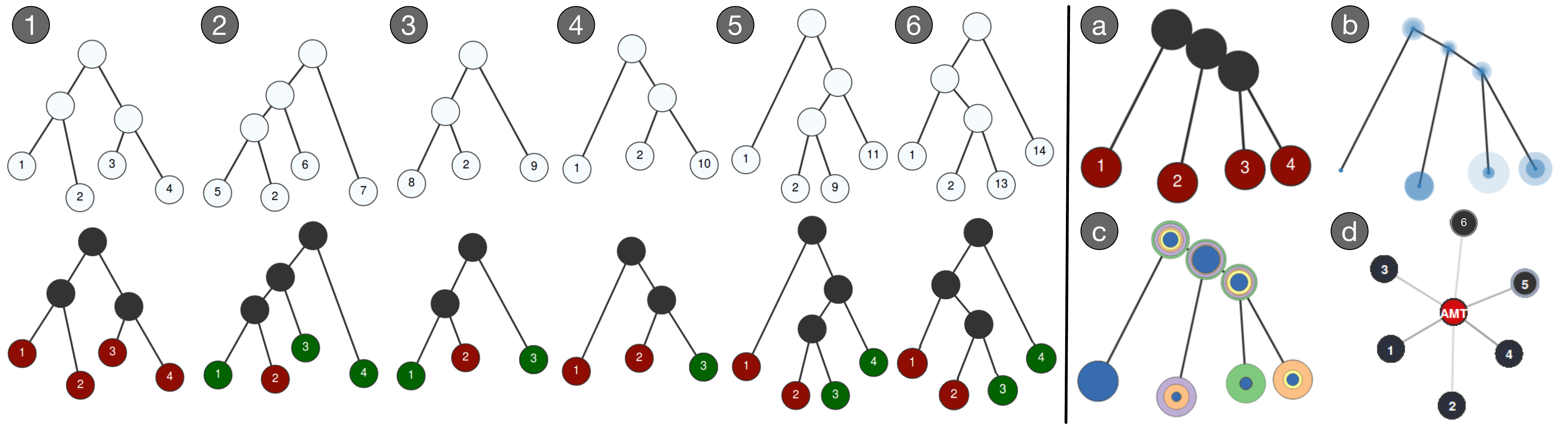}
  \vspace{-6mm}
  \caption{Computing a 1-center tree as a structural average for an input ensemble of six leaf-labeled merge trees in partial agreement. 
  Top row on the left: six input leaf-labeled merge trees, each visualized by a node-link diagram. 
  Bottom row on the left: each input tree is visualized with original labels in red and updated labels in green.
  Right: various uncertainty visualizations for the 1-center tree: (a) 1-center tree with leaf labels; (b) variational vertex consistency plot; (c) statistical vertex consistency plot.
  (d): summary plot that shows the interleaving distance between each input tree and the 1-center tree (denoted as AMT).}
   \label{fig:usage-partial}
}

\abstract{Physical phenomena in science and engineering are frequently modeled using scalar fields.
In scalar field topology, graph-based topological descriptors such as merge trees, contour trees, and Reeb graphs are commonly used to characterize topological changes in the (sub)level sets of scalar fields.
One of the biggest challenges and opportunities to advance topology-based visualization is to understand and incorporate uncertainty into such topological descriptors to effectively reason about their underlying data.
In this paper, we study a structural average of a set of labeled merge trees and use it to encode uncertainty in data.
Specifically, we compute a 1-center tree that minimizes its maximum distance to any other tree in the set under a well-defined metric called the interleaving distance.
We provide heuristic strategies that compute structural averages of  merge trees whose labels do not fully agree.
We further provide an interactive visualization system that resembles a numerical calculator that takes as input a set of merge trees and outputs a tree as their structural average.
We also highlight structural similarities between the input and the average and incorporate uncertainty information for visual exploration.
We develop a novel measure of uncertainty, referred to as consistency, via a metric-space view of the input trees.
Finally, we demonstrate an application of our framework through merge trees that arise from ensembles of scalar fields.
Our work is the first to employ interleaving distances and consistency to study a global, mathematically rigorous, structural average of merge trees in the context of uncertainty visualization.
} 


\CCScatlist{
  Topological data analysis, uncertainty visualization, merge trees
}

\begin{document}
\maketitle


\section{Introduction}
\label{sec:introduction}

In topological data analysis and visualization, topological descriptors have been used to understand and summarize the shape of complex data in science and engineering, ranging from gene expression of breast cancer tumors~\cite{LumSinghLehman2013} to high-throughput screening of nanoporous materials~\cite{LeeBarthelDlotko2018}.
For data modeled as scalar fields, the most popular descriptors include vector-based such as persistence diagrams~\cite{EdelsbrunnerLetscherZomorodian2002} and barcodes~\cite{Ghrist2008, CarlssonZomorodianCollins2004}, graph-based such as merge trees~\cite{BeketayevYeliussizovMorozov2014}, contour trees~\cite{CarrSnoeyinkAxen2003}, and Reeb graphs~\cite{Reeb1946}, as well as complex-based such as  Morse and Morse-Smale complexes~\cite{GerberPotter2012,EdelsbrunnerHarerZomorodian2003,EdelsbrunnerHarerNatarajan2003}.

These topological descriptors provide meaningful abstractions, reduce the amount of data to be processed, utilize sophisticated hierarchical representations that capture features at multiple scales, and enable progressive feature simplifications.
However, as pointed out by Heine et al.~\cite{HeineLeitteHeike2016}, one of the biggest challenges and opportunities is to develop original approaches that \emph{incorporate uncertainty} into topological descriptors  to advance topology-based visualization.
In this paper, we focus on \emph{merge trees}, which are a special type of topological descriptor that tracks the evolution of connected components in the sublevel sets of scalar fields~\cite{MorozovBeketayevWeber2013}.
We compute \emph{structural averages} of a set of labeled merge trees (referred to as an \emph{ensemble}), and utilize such averages in uncertainty visualization.
Our work is motivated from a statistical and a visualization perspective.

\para{Statistics on topological descriptors.}
In statistics, the concept of an \emph{average} refers to a measure of central tendency.
Given a set of numbers, the notions of \emph{mean}, \emph{median}, and \emph{mode} are considered as \emph{numeric averages} in different contexts.
In this paper, we study a certain \emph{structural average} among a set of labeled merge trees. In particular, we would like to find a \emph{metric 1-center} of a finite set of labeled merge trees; that is, an \emph{average tree} that minimizes the maximum distance to any other tree in the set.
Given a metric space $(X, d)$, a $1$-center   of a finite point set $P = \{p_1, \cdots, p_m\} \subset X$ is a point $c \in X$ which satisfies
$c =\displaystyle \arg\min_{x \in X} \max_{p \in P} d(x, p),$
i.e.,~$c  \in X$ is the center of a minimum enclosing the ball of $P$.
Here, our metric space is the space of all labeled merge trees equipped with a metric referred to as the \emph{interleaving distance for labeled merge trees} \cite{MunchStefanou2018}. The $1$-center is our notion of an average tree.

Our implementation in computing a 1-center is a first step toward performing statistical analysis of merge trees.
By introducing a potentially extendable metric between labeled merge trees, we envision a complete suite of statistics describing the probability distribution of graph-based topological descriptors.

\para{Visualization of uncertainty.}
Advances in technology such as increased bandwidth, storage, and  computational power have led to a large amount of complex data.
To effectively and accurately communicate such data to scientists via visualization, we should pay close attention to information about uncertainty, including accuracy, confidence, variability, and model bias~\cite{BonneauHegeJohnson2014}.

Conveying uncertainty information through visualization, i.e., \emph{uncertainty visualization}, is a very active area of research; e.g., see surveys~\cite{GrietheSchumann2006,ThomsonHetzlerMacEachren2005,PotterRosenJohnson2012}.
Uncertainty information is often summarized via statistical quantities (such as mean, median, and standard deviation) and encoded with data via color, opacity, texture, glyphs, animation, etc.~\cite{PotterRosenJohnson2012}.
Uncertainty for 1D scalar fields is often expressed \emph{point-wise} as error bars or boxplots~\cite{Potter2006}.
In the case of 2D and 3D scalar fields, uncertainty can be encoded \emph{feature-wise} within spatial domains  (such as contours~\cite{PothkowHege2011,PothkowWeberHege2011,WhitakerMirzargarKirby2013}, 2D surfaces~\cite{PotterKirbyXiu2012}, and 3D volumes~\cite{PotterKrugerJohnson2008,DjurcilovKimLermusiaux2001}), as well as \emph{structure-wise} as trees~\cite{LeeRobertsonCzerwinski2007} or lattice graphs~\cite{CollinsCarpendalePenn2007}.

The challenges associated with the study of structure-wise uncertainty in graph-based descriptors can be attributed, in part, to the difficulty in characterizing and quantifying their structural uncertainty.
In particular, ensemble datasets combining multiple realizations of a phenomenon are often used to mitigate the effects of uncertainty.
State-of-the-art approaches typically aggregate over inherent data dimensions to reduce the structure and size of the data.
Unfortunately, these aggregations can lose the richness in both global and  local structures.
For instance, a scalar field obtained by averaging the values from all the ensemble scalar field members at each data point (as a form of aggregation) was treated previously as the \emph{ensemble mean}, and the contour tree of the ensemble mean was referred to as the \emph{mean contour tree}~\cite{WuZhang2013}.

In this regard, our work differs from previous approaches significantly.
First, instead of computing a merge tree from an average scalar field, we calculate an average merge tree \emph{directly} from a set of input trees that perhaps arise from an ensemble, and use such an  average to encode uncertainty.
Second, constructing an average labeled merge tree has a clean, mathematical foundation via a metric-space view, and does not rely on ad hoc operations.
To the best of our knowledge, our work is the first to employ interleaving distances and consistency measures to study a mathematically rigorous  structural average of merge trees in the context of uncertainty visualization.

\para{Contributions.}
In this paper, we compute and visualize a structural average of labeled merge trees. Our work builds upon theoretical foundations regarding interleaving distances between labeled merge trees~\cite{GasparovicMunchOudot2019}.
Such a distance is chosen because, unlike many of the other metrics available for merge trees, it is easily computable.
It is also stable with respect to the input function data~\cite{MunchStefanou2018}, and has a well-defined 1-center as well as geodesics.
Our main contributions are the algorithms, implementations, and visualization design in moving from theory to practice, all of which are highly nontrivial.
\begin{itemize}\denselist
\item We provide an interactive visualization system to demonstrate the utilities of our proposed algorithms. It resembles a numerical calculator that takes as input an ensemble of leaf-labeled merge trees and outputs a tree as their structural average. 
\item We introduce heuristic strategies that complete the labelings for an ensemble of merge trees whose labels do not fully agree in order to provide structural averages.
\item We highlight structural similarities between the input and the average tree and incorporate uncertainty information for visual exploration. To achieve this, we develop a novel measure of uncertainty for each vertex in the tree, via a metric-space view of the input trees. This measure is also flexible, allowing a local-global tradeoff in understanding structure variations.
\item We demonstrate an application of our framework through merge trees that arise from ensembles of scalar fields.
\end{itemize}
Our framework is applicable to merge trees of the most general form; we give an additional example of applying our framework to merge trees derived from neuron morphology in Appx.~\ref{appendix:results}. 
Our visualization tool and algorithms are released open source under MIT license on Github: \url{https://github.com/tdavislab/amt}. 

\section{Related Work}
\label{sec:relatedwork}

We review the most relevant literature on uncertainty visualization of scalar fields, with a focus on graph-based topological descriptors and topological features.
Our work is primarily concerned with structure-wise data uncertainty of scalar fields; for vector and tensor fields, see an overview in~\cite{HeineLeitteHeike2016}.

\para{Graph-based topological descriptors.}
Graph-based topological descriptors include
merge trees~\cite{BeketayevYeliussizovMorozov2014}
(also known as barrier trees~\cite{FlammHofackerStadler2002} or join  trees~\cite{CarrSnoeyinkAxen2003}),
contour trees~\cite{CarrSnoeyinkAxen2003},
Reeb graphs~\cite{Reeb1946},
mapper graphs~\cite{SinghMemoliCarlsson2007},
and joint contour nets~\cite{CarrDuke2013}.
These descriptors are graph-based representations to illustrate how the topology of level sets or sublevel sets of scalar fields changes with a scalar value parameter.

For a topological space $\Xspace$ equipped with a function $f: \Xspace \to \Rspace$, the merge tree encodes the connected components of the sublevel sets $f^{-1}(-\infty,a]$ for $a \in \Rspace$.
A closely related descriptor, the Reeb graph, encodes the connected components of the level sets $f^{-1}(a)$ instead.
The contour tree~\cite{CarrSnoeyinkAxen2003} is a type of Reeb graph when $\Xspace$ is simply connected.
All these descriptors are related to Morse theory~\cite{Milnor63} and level-set topology through relations among critical points.
They are widely applied in scientific visualization~\cite{WeberBremerPascucci2007,OesterlingHeineJaenicke2011,WidanagamaachchiJacquesWang2017}.

Since a contour tree of a function $f: \Xspace \to \Rspace$ can be constructed by carefully combining the merge trees of $f$ and $-f$ in linear time~\cite{CarrSnoeyinkAxen2003}, merge tree visualization shares the same design space as that of a contour tree.
Contour trees are often visualized with node-link diagrams in two or three dimensions~\cite{PascucciCole-McLaughlin2002,CarrSnoeyinkAxen2003,TakahashiTakeshimaFujishiro2004,PascucciMcLaughlinScorzelli2004}.
Such diagrams are simple and powerful tools for abstract data representations~\cite{WeberBremerPascucci2007}, contour extraction~\cite{KreveldOostrumBajaj1997}, and data explorations in various application domains \cite{PascucciMcLaughlinScorzelli2004,BajajPascucciSchikore1997}.
Many attempts have been made in contour tree visualization to overcome difficulties in visual interpretation, visual clutter, and missing topological features~\cite{WuZhang2015}.

\para{Uncertainty visualization of topological features.}
Critical points and contours (level sets, iso-surfaces) are important features for the study of scalar field topology.
When data is affected by uncertainty, visualization of such topological features should adapt accordingly.

Mihai and Westermann~\cite{MihaiWestermann2014} measure the likelihood of the occurrence of critical points with respect to both the positions and types of the critical points. Specifically, when the data uncertainty is described by a Gaussian distribution, confidence intervals are derived for the gradient and the determinant and trace of the Hessian matrix in scalar field ensembles to infer confidence regions for critical points~\cite{MihaiWestermann2014}.
Gunther et al.~\cite{GuntherSalmonTierny2014} characterize critical points and their spatial relation for 2D uncertain scalar fields, where each vertex in a regular grid is assigned a probability density function (PDF) describing its scalar value.
They identify so-called mandatory critical points -- spatial regions and function ranges where critical points have to occur in any realization of the input based on the PDF.

To visualize the effect of uncertainty on contours, envelopes within a gridded domain are extracted to indicate in which volume the contour will lie (with a certain confidence)~\cite{ZehnerWatanabeKolditz2010,PangWittenbrinkLodha1997}.
Uncertainty associated with a contour can also be rendered via animation~\cite{Brown2004} or as a collection of points where each point is displayed from ``its original location along the surface normal by an amount proportional to the uncertainty at that point"~\cite{GrigoryanRheingans2004}.
Positional and geometrical variations of contours are captured by variability in gradients for uncertain scalar fields~\cite{PfaffelmoserMihaiWestermann2013}.
Positional uncertainty of contours could also be encoded by spatial correlation~\cite{PfaffelmoserWestermann2012,PfaffelmoserWestermann2013} or numerical sensitivity~\cite{PothkowHege2011}.
Building on the notions of functional boxplots and data depth, contour boxplots~\cite{WhitakerMirzargarKirby2013} display statistical quantities analogous to the mean, median, and order statistics for ensembles of contours.

Finally, from an algorithmic perspective, probabilistic marching cubes~\cite{PothkowWeberHege2011} and positionally uncertain iso-contours~\cite{PothkowHege2011} study the uncertainties inherent in computing the visual representations.

\para{Uncertainty visualization of graph-based decriptors.}
Lee et al.~have introduced CandidTree, which merges two trees into one and visualizes both location and subtree structural uncertainty~\cite{LeeRobertsonCzerwinski2007}.
Wu et al. have developed an interactive visualization tool that uses contour trees as abstract data representations to explore data-level uncertainty, contour and topology variability~\cite{WuZhang2013}.
Kraus has employed grayscale morphology to visualize uncertain substructures in contour trees~\cite{Kraus2010}.
Zhang et al.~have proposed sampling-based Monte Carlo methods to study contour trees of uncertain terrains, where uncertainty lies in the height function described by a probability distribution~\cite{ZhangAgarwalMukherjee2015}.
The work most relevant to ours is~\cite{WuZhang2013}, where a mean contour tree is computed as the contour tree of the mean of an ensemble.
However, our work differs significantly from~\cite{WuZhang2013} in that instead of computing a tree from an average of the ensemble members, we compute an average tree directly from a set of input trees (that potentially arise from ensemble members).

\para{Distances between topological structures.}
Recently, many metrics have been proposed for merge trees, often by way of a restriction from a metric on the more general Reeb graph \cite{BeketayevYeliussizovMorozov2014,MorozovBeketayevWeber2013,SilvaMunchPatel2016,Bauer2014, Bauer2015b,BauerDiFabioLandi2016,BauerLandiMemoli2017,SridharamurthyMasoodKamakshidasan2018, CarriereOudot2017}.
In this paper, we focus on the interleaving distance for labeled merge trees \cite{MunchStefanou2018}.
This distance is an example of an interleaving distance between persistence modules \cite{Chazal2009b}, which is brought to graph-based descriptors such as merge trees \cite{MorozovBeketayevWeber2013} and Reeb graphs \cite{SilvaMunchPatel2016} via category theory \cite{BubenikSilvaScott2014,SilvaMunchStefanou2018}.

The application of the interleaving distance to labeled merge trees can also be viewed as an interpretation of a metric for phylogenetic trees \cite{CardonaMirRossello2013}.
Our framework differs from the previous work as it relies on a clean and simple metric-space view of input trees that is equipped with geodesics, as well as easy-to-implement algorithms.

\section{Technical Background}
\label{sec:background}

\begin{figure*}
\centering
\includegraphics[width=2.05\columnwidth]{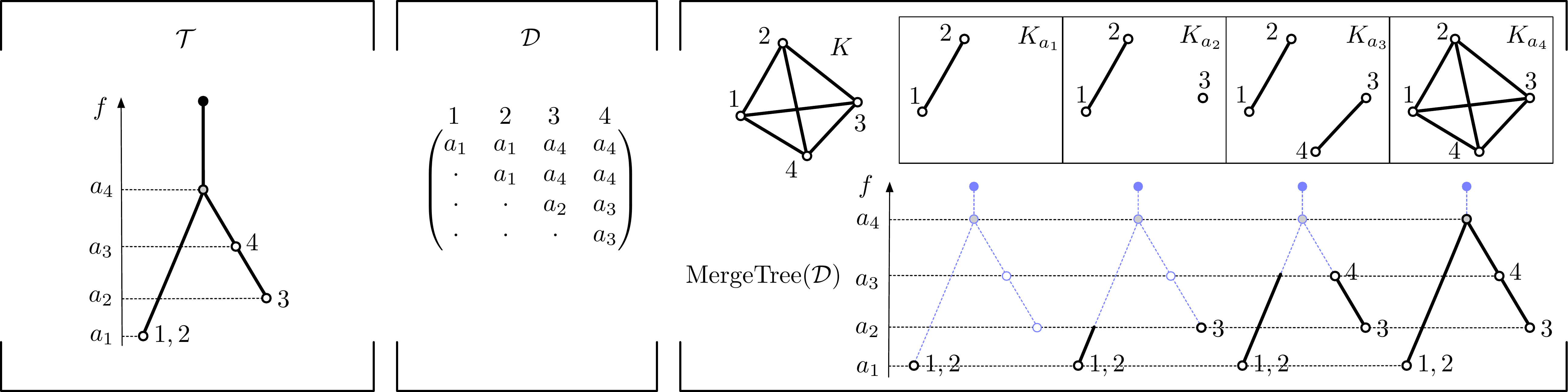}
\vspace{-3mm}
\caption{An example of a labeled merge tree (left) that permits labels on internal vertices and multiple labels on a single leaf, together with its induced matrix (middle). The solid black vertex is the root, and white vertices are labeled. The tree has four nonroot vertices and two leaves.
Right: construction of a labeled merge tree from the ultra matrix that recovers the original tree.}
\vspace{-3mm}
\label{fig:lmt-combined}
\end{figure*}

We first review mathematical notions in graph theory and computational topology, including merge trees, labeled merge trees, and the interleaving distance on labeled merge trees.
We then introduce leaf-labeled merge trees with full, partial, or no label agreements, which are structures we deal with in our algorithms.
We end this section by giving an example of a \emph{scalar field induced merge tree}, that is, a merge tree that arises from a scalar function on a topological space.

\subsection{Theoretical Foundations for Labeled Merge Trees}

Our implementation and visualization design is built upon theoretical foundations established in~\cite{GasparovicMunchOudot2019}, which focus on labeled merge trees of the most general form, and which we review below.

\para{Merge trees and labeled merge trees.}
A \emph{tree} is an undirected graph in which any two vertices (nodes) are connected by a unique path.
A \emph{rooted tree} is a tree in which a special vertex is chosen to be the \emph{root}.
The \emph{degree} of a vertex is the number of edges incident to the vertex.
We assume the root is of degree $1$.

Let $T = (V, E)$ denote a rooted tree with vertex set $V$ and edge set $E$.
The \emph{leaves} $L \subset V$ of $T$ are the nonroot vertices of degree $1$; other nonroot vertices are \emph{internal vertices}.

\begin{definition}
\label{def:mt}
 A \emph{merge tree} is a pair $(T, f)$ consisting of a rooted tree $T$ and a function $f:V \to \Rspace \cup \{\infty\}$ such that $f(u) \neq f(v)$ for all $uv \in E$,  $f(v) = \infty$ if and only if $v$ is the root, and every nonroot vertex has exactly one neighbor with a higher  function value.
\end{definition}

A scalar field induced merge tree is a special case of a merge tree in Definition~\ref{def:mt}, which we review in Sec.~\ref{subsec:scalar-mt}.
We require $f(v) = \infty$ for the root $v$ for technical reasons. We also require $f(u) \neq f(v)$ for all $uv \in T$; in practice, this can be achieved by the simulation of simplicity~\cite{EdelsbrunnerMucke1990}.
  For any pair of vertices $u, v \in V$, we write $\LCA(u,v) \in V$ for their \emph{lowest common ancestor}; that is, the unique vertex of minimum function value such that the unique path from $\LCA(u,v)$ to either $u$ or $v$ strictly decreases in the value of $f$.
Then $f(a(u,v))$ denotes its function value.
For ease of notation, let $[n] := \{1,\ldots,n\}$ denote a label set.

 \begin{definition}
 \label{def:lmt}
 A \emph{labeled merge tree} $\T$ is a triple $(T, f, \mu)$ that consists of a merge tree $(T, f)$ along with a map $\mu:[n] \to V$ called a \emph{labeling} that is  surjective on the set of leaves $L \subset V$.
 When we need to emphasize the number of labels, we will call this an $n$-labeled merge tree.
 \end{definition}

Let $|T| = m$ be the size of the merge tree, or in other words, the number of nonroot vertices in the tree.
Note that the definition permits trees with labeled internal vertices as well as vertices with multiple labels; see Fig.~\ref{fig:lmt-combined} (left) for an example.
Surjectivity on $L$ means that $n$, an arbitrary positive integer, satisfies $n \geq |L|$.
For practical purposes, we will usually have $n \leq m$, but this is by no means required.
Finally, a merge tree $(T,f)$ without a labeling is referred to as an \emph{unlabeled merge tree}.

\para{Induced matrix and interleaving distance.}
We can build a matrix from a labeled merge tree.
The \emph{induced matrix} of an $n$-labeled merge tree $\T=(T, f, \mu)$ is the symmetric matrix $\IM \in \Rspace^{n \times n}$ where
$\IM_{ij} = f( \LCA(\mu(i),  \mu(j)))$.
See Fig.~\ref{fig:lmt-combined} (middle) for an example of an induced matrix.
The reason for turning the trees into matrices is that we can immediately use a natural choice of distance between the matrices to obtain a distance between the trees as follows:

\begin{definition}
\label{def:interleaving}
Given two $n$-labeled merge trees $\T^1=(T_1, f_1, \mu_1)$ and $\T^2=(T_2, f_2, \mu_2)$, the \emph{(labeled) interleaving distance} between them is defined to be the $L_\infty$ distance between their corresponding induced matrices $\IM^1$ and $\IM^2$, namely
$d_I(\T^1, \T^2) = ||\IM^1 - \IM^2||_{\infty}.$
\end{definition}

Notice that Definition~\ref{def:interleaving} requires that the two given trees have the same labeling, so that they are both $n$-labeled merge trees.

\para{Labeled merge tree of a valid matrix.}
We can also turn certain matrices back into trees.
A symmetric matrix $\IM \in \Rspace^{n \times n}$ is called \emph{valid} if $\IM_{ii} \leq \IM_{ij} $ for all $i,j$.
A valid matrix  encodes a function $f$ on the complete graph $K$ of $n$ vertices modeled after $n$ labels, with function value $\IM_{ii}$ on the vertex $i$ and $\IM_{ij}$ on the edge $ij$.

The \emph{labeled merge tree of a valid matrix} $\IM \in \Rspace^{n \times n}$, denoted $\MT(\IM)$, is the labeled merge tree of the complete graph with the induced function $f$.
For a fixed $a \in \Rspace$, let $K_a = f^{-1}(-\infty, a]$.
Such a merge tree is obtained by tracking the connected components of $K_a$ as $a$ increases.
See Fig.~\ref{fig:lmt-combined} (right) for an example and Sec.~\ref{sec:implementation} for implementation details.

A valid matrix $\IM$ is called \emph{ultra} if
$\IM_{ij} \leq \max \{\IM_{ik}, \IM_{kj}\}$.
The induced matrix of a labeled merge tree is an ultra matrix; in fact, this construction induces a bijection between the space of labeled merge trees and the space of ultra matrices~\cite{GasparovicMunchOudot2019}.

It has been shown that the interleaving distance defined above enjoys nice  properties including stability~\cite{MunchStefanou2018}; in particular, two theoretical results from~\cite{GasparovicMunchOudot2019} that are central to our paper are included here for completeness. Theorem~\ref{theorem:1center}, which is a proposition from~\cite{GasparovicMunchOudot2019}, lays the foundation for computing a $1$-center for a set of labeled merge trees.

\begin{theorem}[Proposition LMT 1-Center~\cite{GasparovicMunchOudot2019}]
\label{theorem:1center}
Given $n$-labeled merge trees $\T^1, \ldots, \T^k$ and their corresponding induced ultra matrices $\IM^1, \ldots , \IM^k$,
let $\IM$ be the element-wise $1$-center of $\IM^1, \ldots, \IM^k$.
Set $\T = \MT(\IM)$.
Then $\T$ is a $1$-center of the labeled merge trees $\T^1, ..., \T^k$.
\end{theorem}

Note that when using the $L_\infty$ distance, the 1-center for the matrices is not unique.
However, it is easy enough to compute one of these, which we will make use of in our implementation described in Sec.~\ref{sec:implementation}.
The next theorem implies the existence of an animated morphing between trees, which is also used in our implementation.
\begin{theorem}[Corollary LMT Geodesics~\cite{GasparovicMunchOudot2019}]
\label{theorem:labelintrinsic}
Given any two $n$-labeled merge trees $\T^1$ and $\T^2$ and their corresponding induced ultra matrices $\IM^1$ and $\IM^2$, the family of merge trees
$\left\{
    \T^\lambda:=\MT\left(\IM^\lambda\right)
    \mid{\lambda\in[0,1]}
\right\}$
defines a geodesic between $\T^1$ and $\T^2$ in the metric $d_I$ where
$\IM^\lambda=(1-\lambda)\,\IM^1 + \lambda\, \IM^2$.
\end{theorem}

\subsection{Leaf-Labeled Merge Trees}
We will focus on a restricted class of labeled merge trees defined as follows.
Recall $[n] := \{1,\ldots,n\}$.

\begin{definition}
 \label{def:leaf-lmt}
 A \emph{leaf-labeled merge tree} $\T$ is a triple $(T, f, \omega)$ that consists of a merge tree $(T, f)$ together with a labeling $\omega: S \to L$ that is surjective on the set of leaves $L$, where $S \subseteq [n]$.
 \end{definition}

 Unlike Definition~\ref{def:lmt}, leaf-labeled merge trees allow labels only on the leaves.
 Also, unlike the traditional definition of a \emph{leaf-labeled tree} where each of its leaves is labeled by precisely one element from a given label set~\cite{Johnson2012}, Definition~\ref{def:leaf-lmt} permits multiple labels on the same leaf.
The set of leaf-labeled merge trees can be considered as a subset of labeled merge trees by considering $\omega$ as a restriction of $\mu$ to a subset of $[n]$.
Up to reindexing the label set, the properties involving induced matrices (Theorems~\ref{theorem:1center} and \ref{theorem:labelintrinsic}) and interleaving distance (Definition~\ref{def:interleaving}) therefore apply to leaf-labeled merge trees.

\para{Full agreement, partial agreement and disagreement.}
In our implementation, we work with input data with various levels of labeling, so we discuss some terminology for various types of missing information next.
First, an ensemble of leaf-labeled merge trees $\{(T_i,f_i,\omega_i) \}_{i=1}^k$ is in \emph{full agreement} if the trees all share the same label set; that is, if the domain of $\omega_i$ is the same for all $i$.
Second, an ensemble of leaf-labeled merge trees $\{(T_i,f_i,\omega_i) \}_{i=1}^k$ is in \emph{partial agreement} if the trees are not in full agreement and $\bigcap_{i=1}^k S_i \neq \emptyset$.
Third, an ensemble of leaf-labeled merge trees $\{(T_i,f_i,\omega_i) \}_{i=1}^k$ is in \emph{disagreement} if $\bigcap_{i=1}^k S_i = \emptyset$.
Ensembles of the second and the third type can also be treated as being \emph{partially labeled} and \emph{unlabeled}, respectively, although we rarely use these terminologies here.

\subsection{Merge Trees from Scalar Field Ensembles}
\label{subsec:scalar-mt}
Of particular interest to visualization is merge trees that arise from an ensemble of scalar fields.
Each tree captures the connectivity of the sublevel sets of a scalar field.
In this paper, we refer to it as a \emph{scalar field induced merge tree}.


Given a scalar function $f$ defined on a topological space $\Xspace$, $f: \Xspace \to \Rspace$,
let $\Xspace_a := f^{-1}(-\infty, a]$ denote the sublevel set of $f$ for some $a \in \Rspace$.
Two points $x, y \in \Xspace$ are \emph{equivalent}, $x \sim y$, if they have the same function value, so $f(x) = f(y) = a$, and if they belong to the same component of the sublevel set $\Xspace_a$.
The quotient space $\Xspace$ with respect to the above equivalence relation, $\Xspace / \sim$, is referred to as a scalar field induced merge tree to differentiate it from the more general Definition~\ref{def:mt}. We still refer to such a tree as a merge tree when it is clear from the context.

\begin{figure}[]
\centering
\includegraphics[width=1.00\columnwidth]{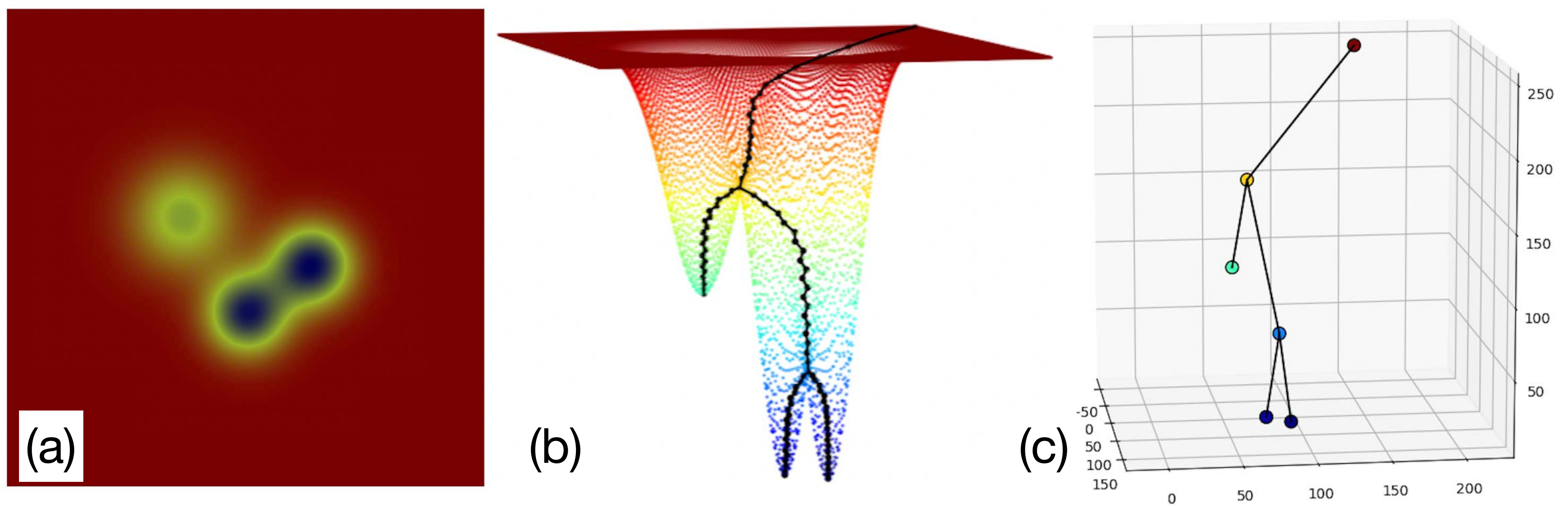}
\vspace{-4mm}
\caption{(a) A 2D scalar field $f$ is generated by a mixture of three Gaussian functions. It is visualized using a rainbow color map: red means high and blue means low values.
(b) The \emph{graph of $f$}, i.e.,~the set of all ordered pairs $(x, f(x))$,  is visualized together with the corresponding augmented merge tree of $f$.
(c) A straight line drawing of an (unaugmented) merge tree of $f$ in 3D.
}
\vspace{-4mm}
\label{fig:mt-example}
\end{figure}

A scalar field induced merge tree is constructed by tracking the evolution of the components of $\Xspace_a$ as we vary the parameter $a$.
Fig.~\ref{fig:mt-example} gives an example: leaves represent the creation of a component at a local minimum, internal vertices (saddles) represent the merging of components, and the root represents the entire space as a single component; see a straight line drawing in Fig.~\ref{fig:mt-example}(c).
We can further augment a scalar field induced merge tree with noncritical points, producing an \emph{augmented} merge tree~\cite[Section 3.3]{CarrSnoeyinkAxen2003}; otherwise it is \emph{unaugmented}.
In this paper, we always visualize an augmented merge tree by embedding its vertices (both critical and noncritical) inside the graph of $f$, as in Fig.~\ref{fig:mt-example}(b).

\subsection{Distance Between Vertices}
\label{subsec:Distances}
Not only do we use distances between the trees themselves, but we also make use of a couple of different distances between vertices of a given merge tree for our implementation.
Such distances impose a metric-space view of input trees and are useful both when handling merge trees with partial or no agreement (Sec.~\ref{subsec:partial-agree}), and in helping to define consistency of vertices in Sec.~\ref{sec:uncertainty}.

Assume we are given a labeled merge tree $\T = (T,f,\omega)$.
The first distance is an intrinsic distance on the merge tree $T$ induced by the function $f$. Specifically, the intrinsic \emph{tree distance between a pair of vertices} $x, y \in V$ is defined to be $d_T(x,y) = |f(x) - f(a(x,y))| + |f(a(x,y)) - f(y)|$.
Note that this is exactly the path length between the vertices if we give every edge a weight equal to the difference in function values of its endpoints.

Now suppose, additionally, that  $\T$ has an embedding $\iota$ in $\Rspace^d$, $\iota: |T| \to \Rspace^d$ (in our experiments, $d=2$ or $3$).
The second distance, the \emph{Euclidean distance between a pair of vertices} $x, y \in V$, is the $L_2$ distance between their embeddings, $d_E(x, y) = ||\iota(x) - \iota(y)||_2$.

\section{Computing 1-Centers of Labeled Merge Trees}
\label{sec:implementation}

Moving from theory to practice, we now discuss implementation details for computing a 1-center of leaf-labeled merge trees.
We compute a structural average under three different scenarios: full agreement, partial agreement, and disagreement. 
For simplicity of explanation, we focus on leaf-labeled merge trees.
Our algorithms and uncertainty encodings (Sec.~\ref{sec:uncertainty}) can be easily adapted to handle general merge trees with labeled internal vertices (Definition~\ref{def:lmt}).

\subsection{Full Agreement}
\label{subsec:full-agree}
We start with the simplest case: given an ensemble of leaf-labeled merge trees in full agreement, we compute its 1-center as a structural average.
As specified by Theorem \ref{theorem:1center}, the main idea is to compute an element-wise 1-center of their induced ultra matrices and convert the resulting valid matrix into a new merge tree.

\begin{figure}[tb]
\centering
\includegraphics[width=0.98\columnwidth]{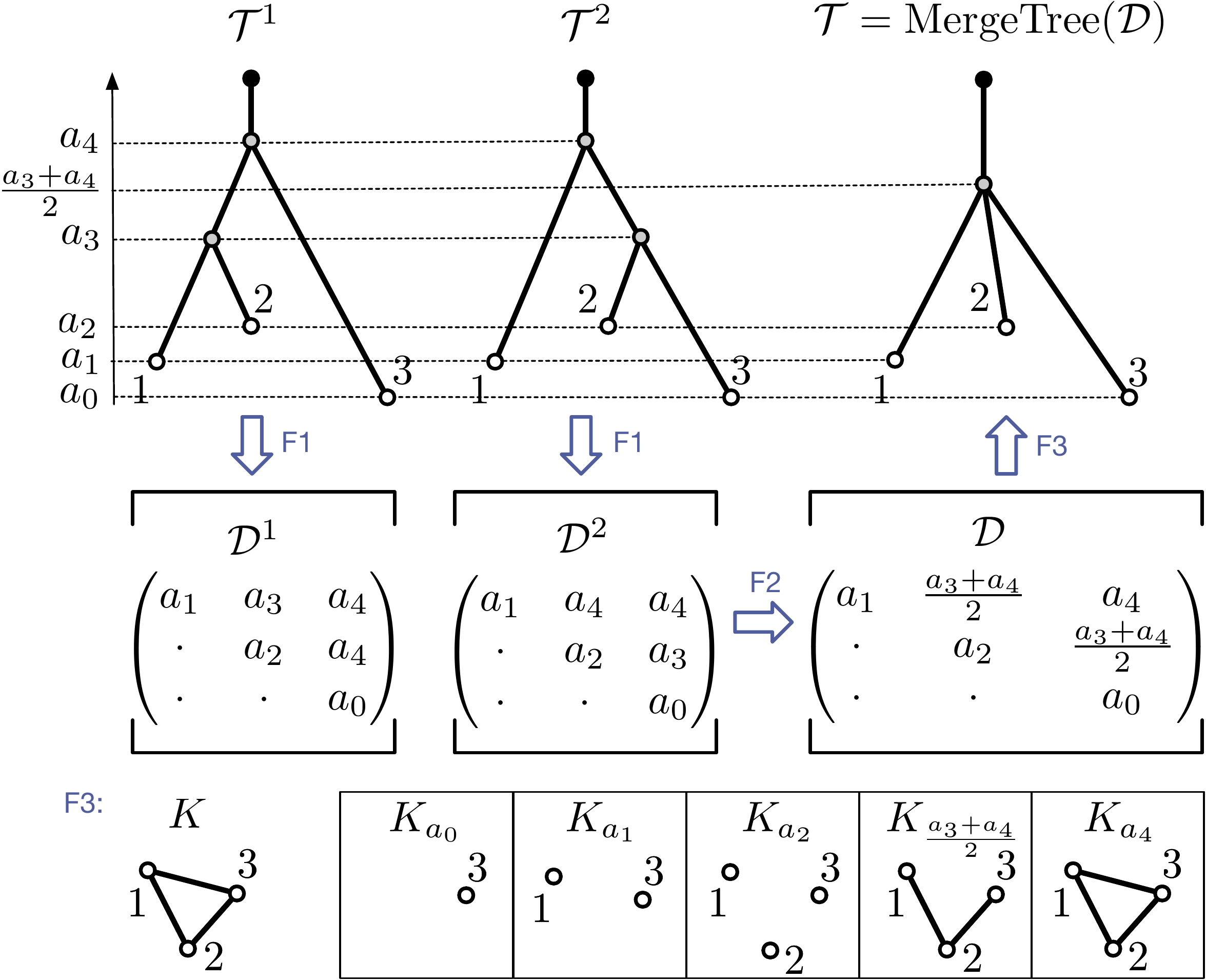}
\caption{Computing a 1-center for a pair of leaf-labeled merge trees in full agreement.
The top row shows two input trees $\T^1$ and $\T^2$ (with three labeled leaves each) together with the output 1-center $\T$.
The middle row displays the induced ultra matrices $\IM^1$ and $\IM^2$ together with their element-wise 1-center, matrix $\IM$, which encodes a function $f$ on the complete graph $K$.
The bottom row shows $K$ and its sublevel set filtration.}
\vspace{-3mm}
\label{fig:lmt-full}
\end{figure}

We start with an ensemble of $k$ leaf-labeled merge trees in full agreement, $\T^1, \cdots, \T^k$, i.e.,~they all share the same label set $S \subseteq [n]$, with $|S| = s$.
On a high level, our algorithm has three simple steps:
\begin{enumerate}\denselist
\item [F1.] Represent each $\T^i$ by its induced ultra matrix $\IM^i$ for $1 \leq i \leq k$.
\item [F2.] Compute the element-wise 1-center $\IM$ of $\IM^1, \cdots, \IM^k$.
\item [F3.] Turn $\IM$ back into a merge tree $\T = \MT(\IM)$.
\end{enumerate}
An example for two leaf-labeled merge trees $\T^1$ and $\T^2$ in full agreement is given in Fig.~\ref{fig:lmt-full}.

Assume  each tree $\T^i = (T_i, f_i, \omega_i)$ induces an ultra matrix $\IM^i \in \Rspace^{s \times s}$ indexed by the shared label set $S \subset [n]$.
For step F2, we construct a choice of 1-center by computing the 1-center of each element,
\begin{equation}
\label{equation:1center}
\IM_{ij} =
    \frac{1}{2} \left|
        \max\{\IM^\ell_{ij}\}_{\ell=1}^k
        -
        \min\{\IM^\ell_{ij}\}_{\ell=1}^k
    \right|.
\end{equation}
Intuitively, $\IM_{ij}$ is the center of a minimum enclosing ball of points $\IM^1_{ij}, \cdots, \IM^k_{ij}$ on a line, i.e.,~the midpoint between the smallest and the largest values along the line.

For Step F3, we turn $\IM$ back into a new merge tree $\T = \MT(\IM)$.
Despite the fact that the matrices $\IM^\ell$ are ultra matrices, the construction of Eq.~\eqref{equation:1center} is not guaranteed to be ultra.
However, it will be a valid matrix, and thus it encodes a function $f$ on the complete graph $K$ among $n$ labels as vertices, $f: K \to \Rspace$. Specifically,
each vertex $v_i \in K$ has value $f(v_i) = \IM_{ii}$; while each edge $e_{ij}=(v_i, v_j) \in K$ has function value $f(e_{ij}) = \IM_{ij}$.
That $\IM$ being a valid matrix guarantees that the function value of an edge is larger than or equal to that of each of its endpoint. 
The returned merge tree follows the standard merge tree construction described below.

Denote the sublevel set of $K$ for $a \in \Rspace$ by $K_a:= f^{-1}(-\infty,a]$.
The merge tree of $f$ is defined on the vertex set of $K$ and keeps track of the connected components in $K_a$.
As $a$ increases from $-\infty$ to $\infty$, vertices in $K$ creates new components in the tree, and edges in $K$ either connect two vertices already from the same components, or merge existing components. We adapt Kruskal's algorithm that runs in time $O(|E| a(|V|))$ ($a$ denotes the inverse of the Ackermann function), processes vertices/edges in a sorted order, and maintains connected components in a disjoint set data structure~\cite{GallerFischer1964, GalilItaliano1991} that supports fast component identification and merging~\cite{SmirnovMorozov2017}.

By Theorem~\ref{theorem:1center}, $\T$ is a 1-center of the leaf-labeled merge trees. Such a 1-center is not necessarily unique; nevertheless it serves as a justifiable structural average as it minimizes the maximum distance to any other tree in the set under $d_I$.
Note that we allow multiple labels on the same leaf to ensure that this definition is well defined.
See an example from Fig.~\ref{fig:usage-full-no} generated from our visualization system.
\begin{figure}[t]
\centering
\includegraphics[width=0.98\columnwidth]{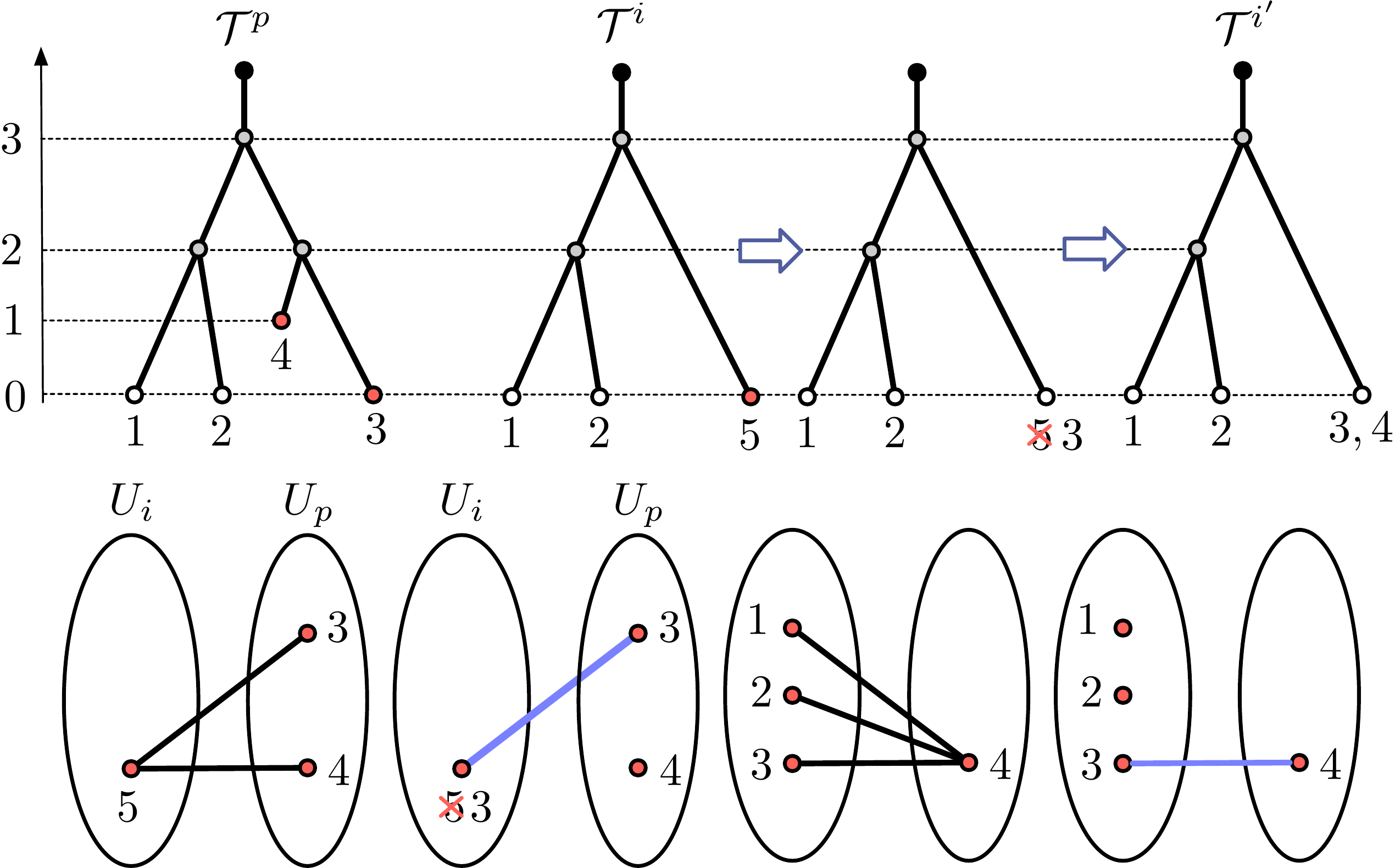}
\caption{Updating the labeling of $\T^i$ against a pivot tree $\T^p$.}
\vspace{-3mm}
\label{fig:lmt-partial-2}
\end{figure}

\begin{figure*}[tb]
\centering
\includegraphics[width=1.98\columnwidth]{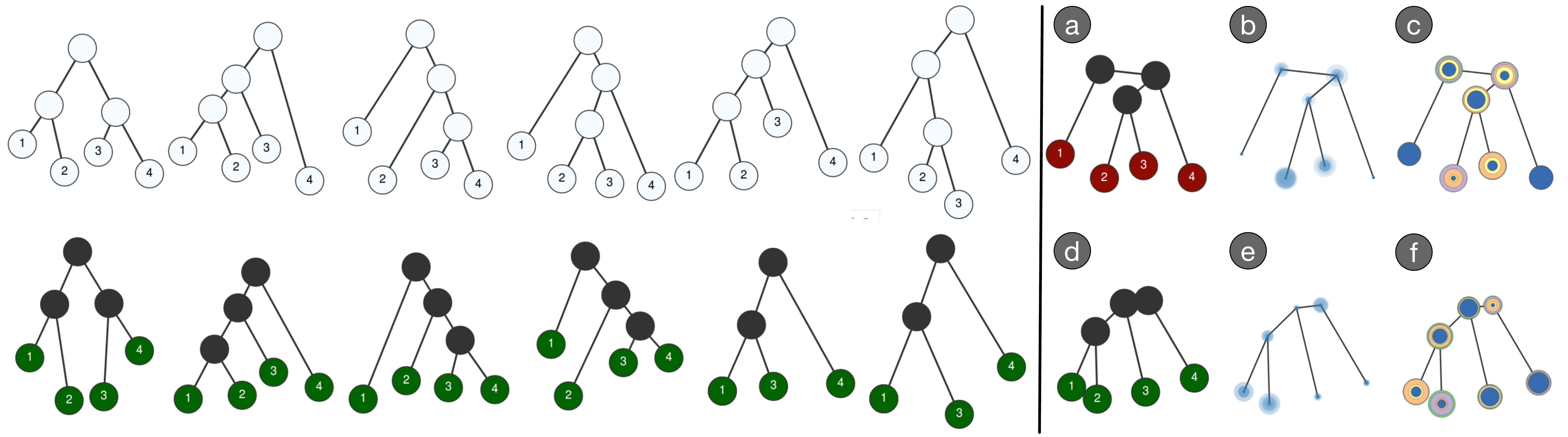}
\vspace{-2mm}
\caption{Computing a 1-center for an ensemble of six leaf-labeled trees in full agreement (top row) and disagreement (botton row). (a, d) 1-center tree;  (b, e) variational vertex consistencies plots; (c, f) statistical vertex consistency plots.
      For the ensemble in disagreement, original labels are omitted; updated labels are green.}
\vspace{-3mm}
\label{fig:usage-full-no}
\end{figure*}

\subsection{Partial Agreement}
\label{subsec:partial-agree}

We now discuss strategies to turn incomplete information in terms of labeling into complete information so that the algorithm of Sec.~\ref{subsec:full-agree} can be utilized.
Given an ensemble of trees whose labels do not fully agree,
finding the best ways to label the remaining vertices boils down to finding the best correspondences under the interleaving distance, which is unfortunately NP-hard~\cite{AgarwalFoxNath2018}.
We thus aim to develop a heuristic approach that is efficient and effective in practice for both the partial agreement and disagreement cases.

Assume we are given an ensemble of leaf-labeled merge trees $\T_1, \cdots, \T_k$ in partial agreement.
If $S_i$ is the label set for tree $\T_i$, let $S = \bigcap_{i=1}^k S_i$ be the shared set of labels.
For simplicity, we assume leaves do not have multiple labels for the input trees.
On a high level, our algorithm is as follows:
\begin{enumerate}\denselist
\item [P1.]
    Select a pivot tree $\T^p = (T_p, f_p, \omega_p)$ with the largest number of leaves among the input trees.
    Let $S_p$ be its label set.
\item [P2.]
    Convert each ensemble member $\T^i = (T_i, f_i, \omega_i)$  to a labeled tree $\T_i'=(T'_i, f_i, \omega'_i)$ by updating its labeling using the label set $S_p$ combined with minimum weight matching.
\item [P3.]
    Compute a $1$-center of the trees $\T_{1'}, \cdots, \T_{k'}$ using Sec.~\ref{subsec:full-agree}.
\end{enumerate}

The key step is P2, for which we first give a pictorial toy example in Fig.~\ref{fig:lmt-partial-2} (top).
The main idea is to study the structural similarities between unmatched leaves by comparing their distances (a linear combination of $d_T$  and $d_E$) to the matched leaves and  solving a minimum weight matching problem.
We also give an example in Fig.~\ref{fig:usage-partial} generated from our visualization system.

\para{Algorithmic details.}
Recall that a \emph{matching} in a bipartite graph is a set of the edges chosen in such a way that no two edges share an endpoint.
A vertex is \emph{matched} if it is an endpoint of one of the edges in the matching. Otherwise it is \emph{unmatched}.
A \emph{maximum matching} is a matching of a maximum number of edges.
An \emph{assignment} (or a \emph{minimum weight matching}) problem, in our setting, is the problem of finding, in a weighted bipartite graph, a maximum matching in which the sum of weights of the edges is as small as possible.

We describe our algorithm with the toy example given in Fig.~\ref{fig:lmt-partial-2}.
Given a pivot tree $\T^p$ with a label set $S_p=\{1,2,3,4\}$, we update the labeling for an ensemble member $\T^i$ using $S_p$.
The matched (shared) label set between the two trees is $S = \{1, 2\}$, and the unmatched label sets in $\T^p$ and $\T^i$ are $U_p = \{3, 4\}$ and $U_i =\{5\}$, respectively.
Our goal is to assign \emph{new} labels to $U_i$ from the set $U_p$ as follows.

First, build $D_p$ to be a pairwise distance matrix with rows corresponding to unmatched labels $U_p$ and columns corresponding to matched labels $S$.
For $x \in U_p$ and $y \in S$, $D_p(x,y)$ is the distance between leaves $\omega_p(x)$ and $\omega_p(y)$.
The matrix $D_i$ is built similarly for tree $\T_i$.
Note that both $D_p$ and $D_i$ have the same number of columns, but potentially a  different number of rows.
In our example, using the tree distance $d_T$, we have
$D_p=
\begin{pmatrix}
  6 & 6 \\
  5 & 5
  \end{pmatrix}$ and 
 $D_i=\begin{pmatrix}
  6 & 6
  \end{pmatrix}.$

Second, construct a complete, weighted bipartite graph with vertex sets $U_i$ and $U_p$ where the weight $c_{xy}$ between a label $x \in U_i$ and $y \in U_p$ is given by the $L_2$ distance between the rows from $D_i$ and $D_p$, respectively.
We find a minimum weight matching in this bipartite graph that gives an assignment of the rows of $D_i$ (unmatched labels in an ensemble member) to the rows of $D_p$ (matched labels in the pivot tree).

Formally, let $b$ be a Boolean matrix where $b_{xy}=1$ if and only if a leaf $x \in U_i$ is assigned to a leaf $y \in U_p$. Then the optimal assignment has a cost: 
$
\min \sum_x \sum_y c_{xy} b_{xy}.
$
We then use the SciPy Python library to find a minimum cost matching, which employs the Hungarian algorithm~\cite{Kuhn1955}.
Since $\T^p$ was chosen to have the maximum number of leaves over all trees, $|U_i| \leq |U_p|$, and thus an optimal assignment will saturate all vertices in $U_i$.
Thus, the output is a complete assignment $\eta: U_i \to U_p$ of minimal cost.
We then define $\omega'_i(x) = \eta(x)$ for $x \in U_i$, and $\omega'_i(x) = \omega_i(x)$ otherwise.

At this point, we might still have unmatched labels $U'_p$ in the pivot tree $\T^p$. 
We use a greedy algorithm to assign each such label to a leaf in the input tree based on its local structure similarity.
In the example, we compute an updated matrix $D'_p$ between the remaining unmatched label $U'_p = \{4\}$ and the matched labels $S' = \{1,2,3\}$, obtaining
$D'_p=\begin{pmatrix}
  5 & 5 & 3
  \end{pmatrix}.$
We then compute the pairwise distance matrix between all matched labels for $\T^i$ as
$
D'_i=
 \begin{pmatrix}
  0 & 2 & 6\\
  2 & 0 & 6\\
  6 & 6 & 0
 \end{pmatrix}.
$
 The weight $c'_{xy}$ between a label $x \in S'_i$ and a label $y \in U'_p$ is again given by the $L_2$ distance between the rows from $D'_i$ and $D'_p$, respectively.
The leaf with the label $x$ now obtains an additional label $y$ when $c'_{xy}$ achieves its minimum.
In the toy example, the leaf in $\T^i$ with a label 3 obtains an additional label 4.
In other words, the labeling $\omega'$ is further updated by including the label $4$ in its domain.

\subsection{Disagreement}
\label{subsec:disagree}

\begin{figure}[tb]
\centering
\includegraphics[width=0.98\columnwidth]{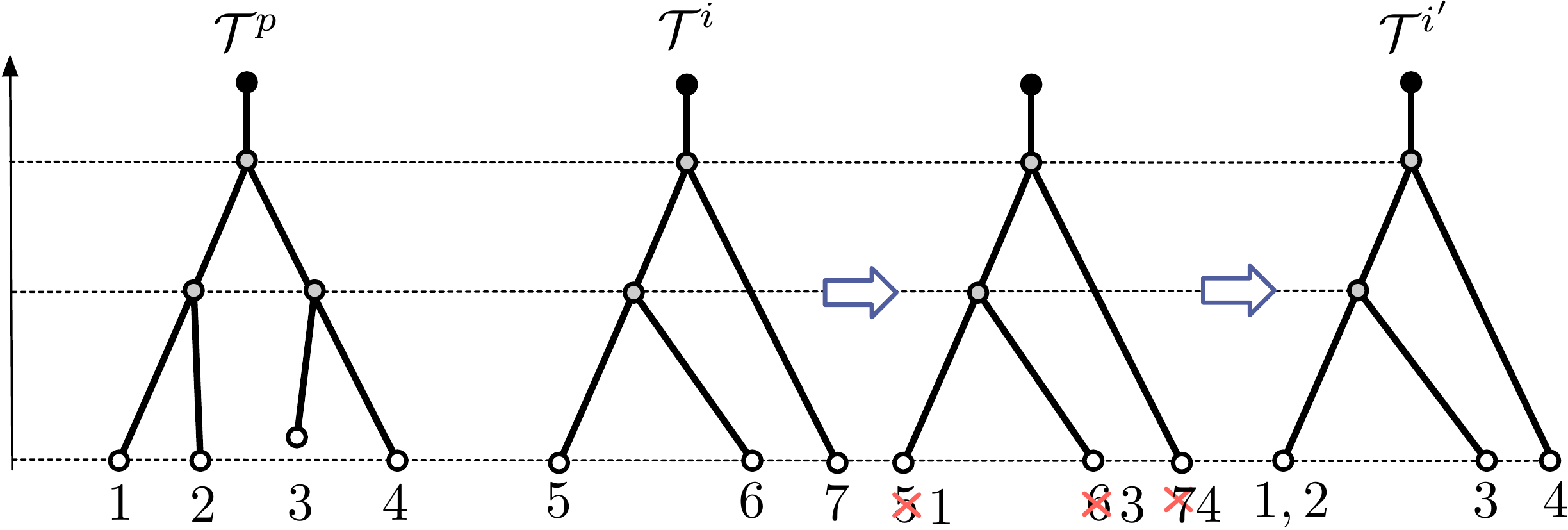}
\vspace{-2mm}
\caption{Updating the labeling of $\T^i$ against a pivot tree $\T^p$.}
\vspace{-3mm}
\label{fig:lmt-disagree}
\end{figure}

To compute a 1-center for an ensemble of leaf-labeled trees in disagreement, we again employ a labeling strategy that converts the trees to leaf-labeled trees in full agreement, and then apply the algorithm in Sec.~\ref{subsec:full-agree}.
To find the labels, however, we now assume each tree $\T_i =(T_i, f_i, \omega_i)$ comes with a geometric embedding $\iota_i: |T_i| \to \Rspace^d$.
Our algorithm below is very similar to the partial agreement case; see Fig.~\ref{fig:lmt-disagree} for an example. 

\begin{itemize}\denselist
\item [D1.] Select a pivot tree $\T^p$ with the largest number of leaves. Let $S_p$ be its label set.
\item [D2.] Update the initial labeling of each ensemble member $\T^i$ by creating a new labeling using the label set $S_p$ and a minimum weight matching. The updated tree is denoted as $\T^{i'}$.
\item [D3.] Compute a $1$-center of the trees $\T^{1'}, \cdots, \T^{k'}$ following Sec.~\ref{subsec:full-agree}.
\end{itemize}

In Step D2, for each ensemble member $\T^i$, we assume $S_i \cap S_p = \emptyset$ (otherwise, we follow the identical algorithm in Sec.~\ref{subsec:partial-agree}).
We create a weighted, complete bipartite graph between the label sets $S_i$ and $S_p$, where the weight $c_{xy}$ between a label $x \in S_i$ and a label $y \in S_p$ is their Euclidean distance in the embedded space, $c_{xy} = d_E(\iota_i(x), \iota_p(y))$.
We again solve a minimum weight matching problem.
The output is a complete assignment $\eta: S_i \to S_p$ of minimal cost.
We then define $\omega'_i(x) = \eta(x)$ for $x \in S_i$.
Since $|S_i| \leq |S_p|$, for any unmatched label $y \in S_p$, we follow a similar strategy as in Sec.~\ref{subsec:partial-agree}, and $\omega'_i$ is updated accordingly.

\section{Encoding Uncertainty}
\label{sec:uncertainty}

Given an ensemble of labeled merge trees $\T^1, \cdots, \T^k$ and their 1-center $\T$, we work toward a visualization that highlights the structural consistency between each  member $\T^i$ and $\T$.
To this end, we develop a novel measure of uncertainty via a metric space view of trees in the ensemble.
This measure is also flexible, allowing a local-global tradeoff in understanding structure variations.
In particular, we compute and visualize \emph{vertex consistency} following a strategy based on Gaussian-weighted cosine similarity.
Our consistency measures apply to general labeled merge trees as in Definition~\ref{def:lmt}, not just leaf-labeled ones.

\begin{figure*}[tb]
\centering
\includegraphics[width=1.75\columnwidth]{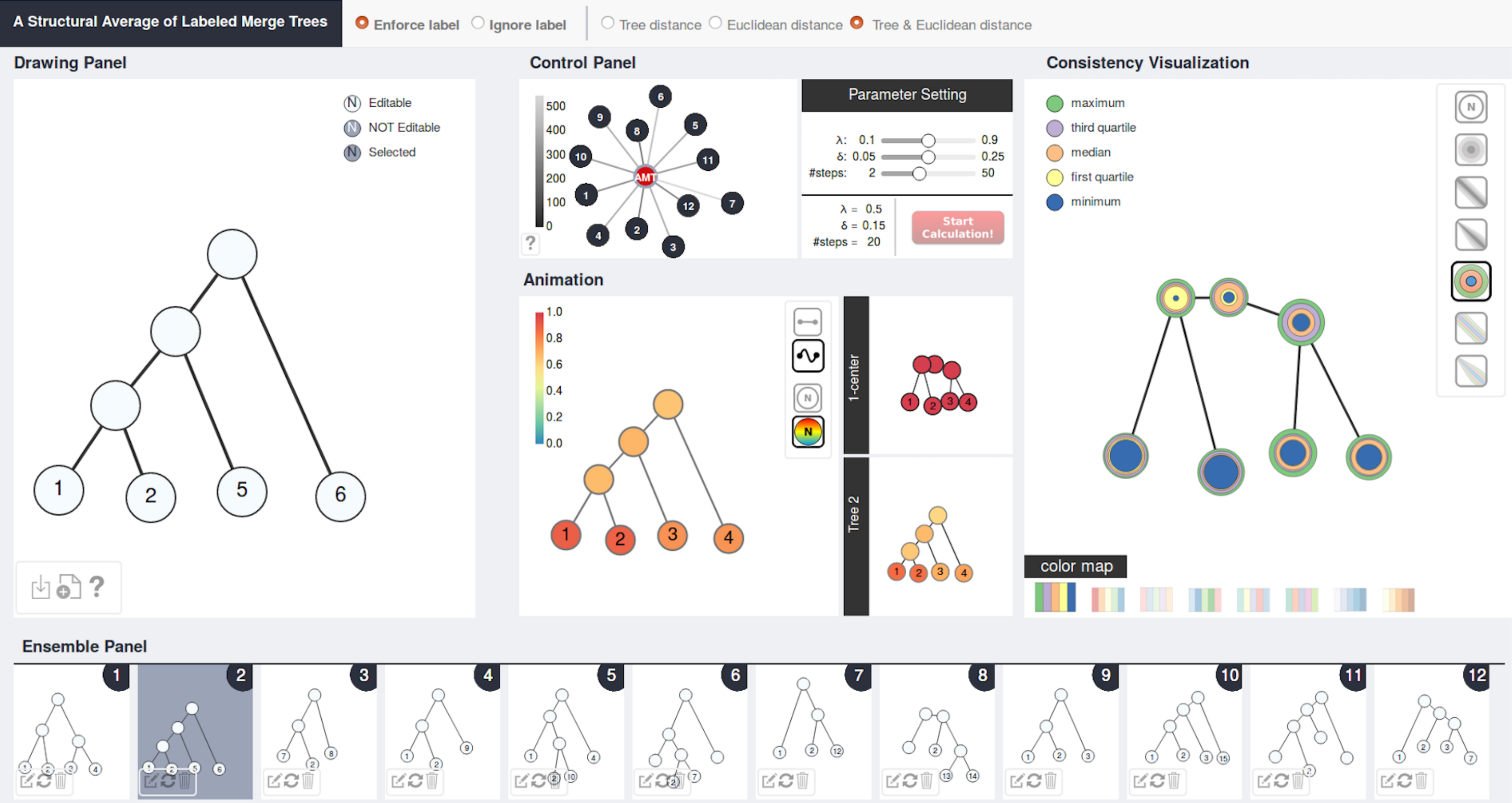}
\vspace{-2mm}
\caption{User interface for the interactive visualization of labeled merge trees and their 1-center.}
\vspace{-3mm}
\label{fig:interface}
\end{figure*}

\subsection{Computing Consistency for a Pair of Trees}

\para{Cosine similarity and weighted cosine similarity.}
 Cosine similarity is a measure of similarity between two nonzero vectors  that measures the cosine of the angle between them.
 Let $A = [A_1, \cdots, A_m] $ and $B = [B_1, \cdots, B_m]$ be two nonzero vectors of length $m$. Their cosine similarity is
$\sm(A, B) = \frac{\sum_i   A_i B_i }
    { \sqrt{\sum_i  A^2_i} \sqrt{\sum_i B^2_i} }.$
With respect to the parameter $\delta$, we use a \emph{Gaussian-weighted cosine similarity} given by
\begin{equation}
\label{equation:wcos}
\sm_\delta(A, B) =
\frac{\sum_i  \left(e^{- \frac{A_i^2 + B_i^2}{\delta^2}} A_i B_i \right) }
    {\sqrt{ \left(\sum_i  e^{- \frac{2A_i^2}{\delta^2}} A^2_i\right)} \cdot \sqrt{\left(\sum_i e^{- \frac{2B_i^2}{\delta^2}} B^2_i\right) }}.
\end{equation}
Note that this similarity provides a ``soft-threshold'' for entries of $A$ and $B$, so that entries smaller than $O(\delta)$  play a more important role in the similarity measure.
In general, smaller values of the $A_i$'s and $B_i$'s are more important.
If we remove all exponential factors in Eq.~(\ref{equation:wcos}), then it becomes the standard cosine similarity.

\para{Vertex consistency.}
Given two labeled merge trees $\T^1 = (T_1, f_1, \mu_1)$ and $\T^2 = (T_2, f_2, \mu_2)$, we want to compute the consistency of vertices in $\T^1$ with respect to vertices in $\T^2$.
For simplicity, we assume that $\mu_1$ and $\mu_2$ are bijective on the set of vertices $V_1$ and $V_2$, respectively (thus all vertices of $\T^1$ and $\T^2$ have unique labels).
Let $|S_1| = |S_2| = s$ be the number of labels.

Let $V_1 = \{ v_1, \ldots, v_s\}$ and $V_2 = \{ w_1, \ldots, w_s\}$ be the labeled vertices for $T_1$ and $T_2$, respectively, where $v_i$ corresponds to $w_i$.
Assume the chosen metrics between vertices in $T_1$ and $T_2$ are $d_1$ and $d_2$, respectively, and
fix a label $l$. 
We wish to measure the ``consistency'' of node $v_l$ of $T_1$ with node $w_l$ of $T_2$. To do this, consider the two vectors $A = [d_1 (v_1, v_l), \ldots, d_1(v_s, v_l)]$ and $B = [d_2(w_1, w_l), \ldots, d_2(w_s, w_l)]$. That is, $A$ (resp. $B$) is the vector of distances from $v_l$ (resp. $w_l$) to all other vertices (ordered by the labels). Intuitively, $A$ (resp. $B$) summarizes how all other vertices relate to $v_l$ (resp. to $w_l$) from a metric point of view. Hence, to measure the consistency between $v_l$ and $w_l$, we use the Gaussian-weighted cosine similarity between these two vectors; that is, the \emph{vertex consistency between $v_l$ and $w_l$} is
$
\sm_{\delta}(v_l, w_l) : = \sm_{\delta}(A,B)
$.
This consistency (similarity) value ranges from $0$ to $1$, and it is $1$ if and only if $A = B$.

Intuitively, $\delta$ is a locality parameter: distances larger than, say, $3\delta$, will essentially be ignored, so this in effect is a soft thresholding where only nearby neighbors of $v_l$ and $w_l$ are being considered.
By adjusting $\delta$, we can change the neighborhood size.
Setting  $\delta = \infty$, we recover the traditional cosine similarity.
The larger $\sm(v_l, w_l)$ is, the more consistent vertices $v_l$ and $w_l$ are with each other.

We define the \emph{vertex consistency} of $\T^1$ (with respect to $\T^2$) as a function defined on the vertices of $\T^1$, $\alpha_1: V_1 \to \Rspace$.
For a vertex with a label $l$, $\alpha_1(v_l) = \sm_{\delta}(v_l, w_l)$ for $v_l \in V_1$ and $w_l \in V_2$.
$\alpha_2: V_2 \to \Rspace$ is defined similarly.

\subsection{Visual Encoding of Consistency for an Ensemble}

We describe visual encodings for vertex consistencies for an ensemble member, variational consistencies, and statistical consistencies for the 1-center tree.

\begin{figure}[tb!]
\centering
\includegraphics[width=0.98\columnwidth]{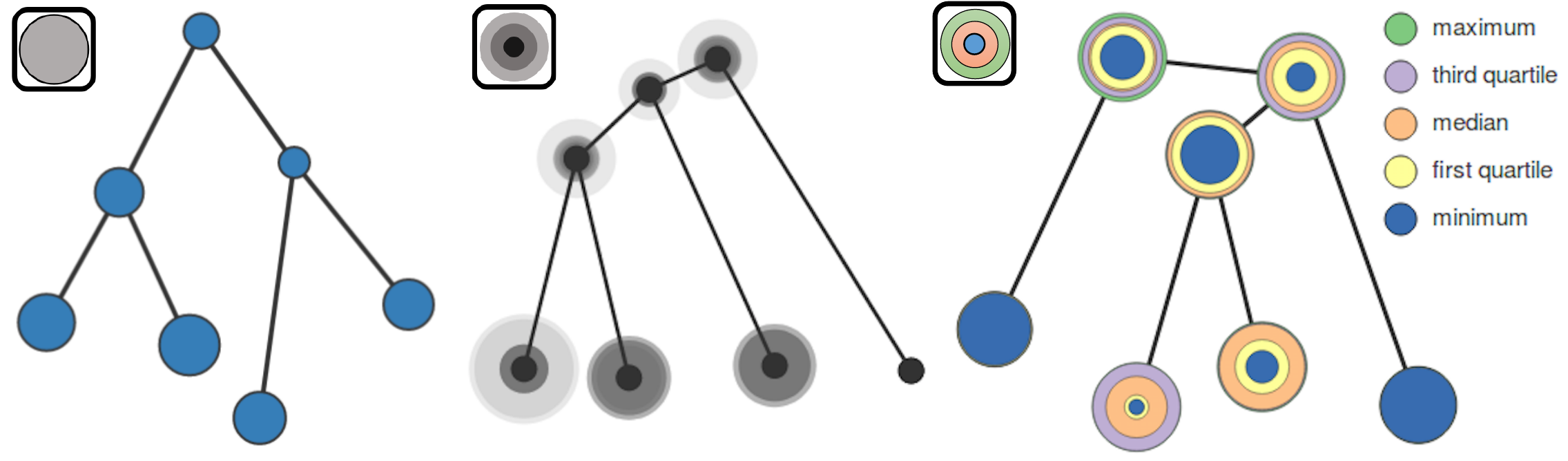}
\vspace{-2mm}
\caption{Left: circular glyphs are used to encode vertex consistencies for an ensemble member. Graduated circular glyphs are used to encode variational (middle) and statistical (right) vertex consistencies for two 1-center trees.}
\vspace{-5mm}
\label{fig:glyphs-vertex}
\end{figure}

We encode vertex consistency for an ensemble member $\T^i$ (with respect to the 1-center $\T$) using glyphs; see Fig.~\ref{fig:glyphs-vertex}(left). 
Specifically, given a vertex consistency $\alpha_i: V_i \to \Rspace$, the radius of each circular glyph at a vertex $v \in V_i$ scales proportional with $\alpha_i(v)$.

We encode variations in vertex consistencies for the 1-center tree $\T$ using visual primitives inspired by~\cite{SanyalZhangBhattacharya2009,SanyalZhangDyer2010}; see Fig.~\ref{fig:glyphs-vertex}(middle).
For the 1-center tree with vertex set $V$, multiple vertex consistency functions are defined with respect to $k$ ensemble members, $\alpha_1,\cdots,\alpha_k: V \to \Rspace$.
Let $\bar{\alpha}$ be their mean value.
Let $\alpha'_i: = |\alpha_i - \bar{\alpha}|$, and we
compute the sequence of variations $\{\alpha'_i,\dots,\alpha'_k\}$ as deviations from the mean $\bar{\alpha}$.
The radius $r_i$ of the $i$-th circular glyph is $r_i = \frac{g \alpha'_i}{2\alpha'}$~\cite{SanyalZhangDyer2010}, where $g$ is the desired spacing between glyphs, and $\alpha'$ is the maximum difference of any $\alpha_i$ to the mean of $\alpha$ in the entire ensemble.
The smaller the glyphs are, the more consistent the ensemble members are with respect to the 1-center.
A small core indicates few outliers whereas a wider core indicates more deviation within members~\cite{SanyalZhangDyer2010}.
In our system, variational consistencies are rendered in a sequential colormap using a single hue.

Inspired by box plots, we visualize the distribution of vertex consistencies at the 1-center $\T$.
For $\alpha_1,\cdots,\alpha_k$ defined at a vertex $v \in V$ of $\T$,
we compute their minimum, first quartile, median, third quartile, and maximum and apply graduated glyphs, as shown in Fig.~\ref{fig:glyphs-vertex}(right).
In our system, statistical consistencies are rendered using a miscellaneous colormap. Finally, the vertex consistency can be extended to edge consistency, see Appx.~\ref{appendix:edge-consistency} for details.

\section{Interactive Visualization and Usage Scenarios}
\label{sec:use-cases}

We provide an interactive visualization system that takes as input an ensemble of leaf-labeled merge trees and outputs a 1-center tree as their structural average.
The system incorporates 1-center computation, animation and uncertainty visualization; see the supplementary video for a demo and Appx.~\ref{appendix:internal} for implementation and Appx.~\ref{appendix:design} for design details.
It is implemented using \textbf{D3.js}, \textbf{Ajax}, \textbf{Flask}, and \textbf{Python}. 

Its user interface is shown in Fig.~\ref{fig:interface}.
The \textbf{drawing panel} allows the user to draw individual merge trees using node-link diagrams and assign initial labels to the vertices.
Each tree is then added to the \textbf{ensemble panel}, where ensemble members can be selected, deleted and edited. 
The \textbf{control panel} provides various options in computing a 1-center tree, whereas an animated sequence between an input tree and the 1-center tree is provided within the \textbf{animation} panel. 
Various consistency measures are visualized in the \textbf{consistency visualization} panel;  see Appx.~\ref{appendix:design} for design details.

We now describe how a user interacts with our visualization system under various usage scenarios.
Thanks to \emph{consistency}, a novel measure of uncertainty for vertices in a tree, our system helps the user  perform tasks that were previously challenging. In particular, we can:
\begin{itemize}\denselist
\item Use the system as a structural calculator:  it takes as input a set of leaf-labeled merge trees and outputs a 1-center tree as their structural average.
\item Perform label diagnostics and correction to reduce data uncertainty and improve structural consistency.
\item Understand structural similarities in a dynamic setting via animations from an ensemble member to the 1-center tree.
\end{itemize}
In addition, our system and its underlying algorithms allow us to study the tradeoff between local and global consistency measures in capturing structural similarities between ensemble members and their structural average, as well as to investigate heuristics in labeling strategies by exploring the tradeoff between intrinsic and extrinsic metrics (see Appx.~\ref{appendix:localglobal} and Appx.~\ref{appendix:inextrinsic} for examples).

\subsection{Computing 1-Centers as Structural Averages}
Suppose we have a numerical calculator. 
A typical usage scenario to obtain an average is to add a set of $k$ input numbers and divide the sum by $k$.
We want to perform similar operations, not with numbers, but with complex structures such as merge trees.

As illustrated in Fig.~\ref{fig:usage-partial}, when a user provides as input an ensemble of six leaf-labeled merge trees in partial agreement 
(top row on the left), the system first applies a (re)labeling strategy to update unmatched labels among the ensemble members while preserving the matched labels (bottom row on the left).
For instance, \emph{Tree 2} has three of its leaves relabeled based on their structural similarities to the pivot \emph{Tree 1}.
 The labeling on the 1-center tree (a) shows leaf correspondences between the input and the output.
The system incorporates uncertainty information on the 1-center tree: variations (b) and distributions (c) in vertex consistencies are encoded by graduated circular glyphs.

We explore the structural variations of the input ensemble via the summary plot (d), where all the ensemble members (in this particular input) have roughly the same interleaving distance to the 1-center (a).
In the variational consistency plot (b), we see that there is a small variation in the vertex consistency measure around leaf 1 for the 1-center as indicated by a tiny blue circular glyph, which means that all six ensemble members are highly consistent in the local neighborhood of leaf 1.
This is also captured by the statistical consistency plot (c) where the minimum, maximum, and medium consistency values coincide at leaf 1.

Using our visualization system, the user can perform computations and uncertainty visualization for trees in full agreement and disagreement in a similar fashion; see Fig.~\ref{fig:usage-full-no}.
For instance, the variational vertex consistency plot in Fig.~\ref{fig:usage-full-no}(b) shows that all six input trees have almost no structural variation at leaves 1 and 4, whereas the distribution of vertex consistencies is much less concentrated at leaves 2 and 3 in the statistical consistency plot Fig.~\ref{fig:usage-full-no}(c), which can be explained in the sense that leaf 1 and leaf 4 in each input tree all have a common lowest ancestor at the root with similar height values. 

In summary, our system enables us to explore the structural variations between input trees and their 1-center by incorporating various means of consistency-based uncertainty visualization; see the supplementary video for a demo.

\subsection{Label Diagnostics and Correction}
Apart from the default functionality in computing a 1-center, perhaps more importantly, we could use our interactive tool to perform label diagnostics and correction to reduce data uncertainty among ensemble members and improve structural consistency.

\begin{figure}[tb]
\centering
\includegraphics[width=0.8\columnwidth]{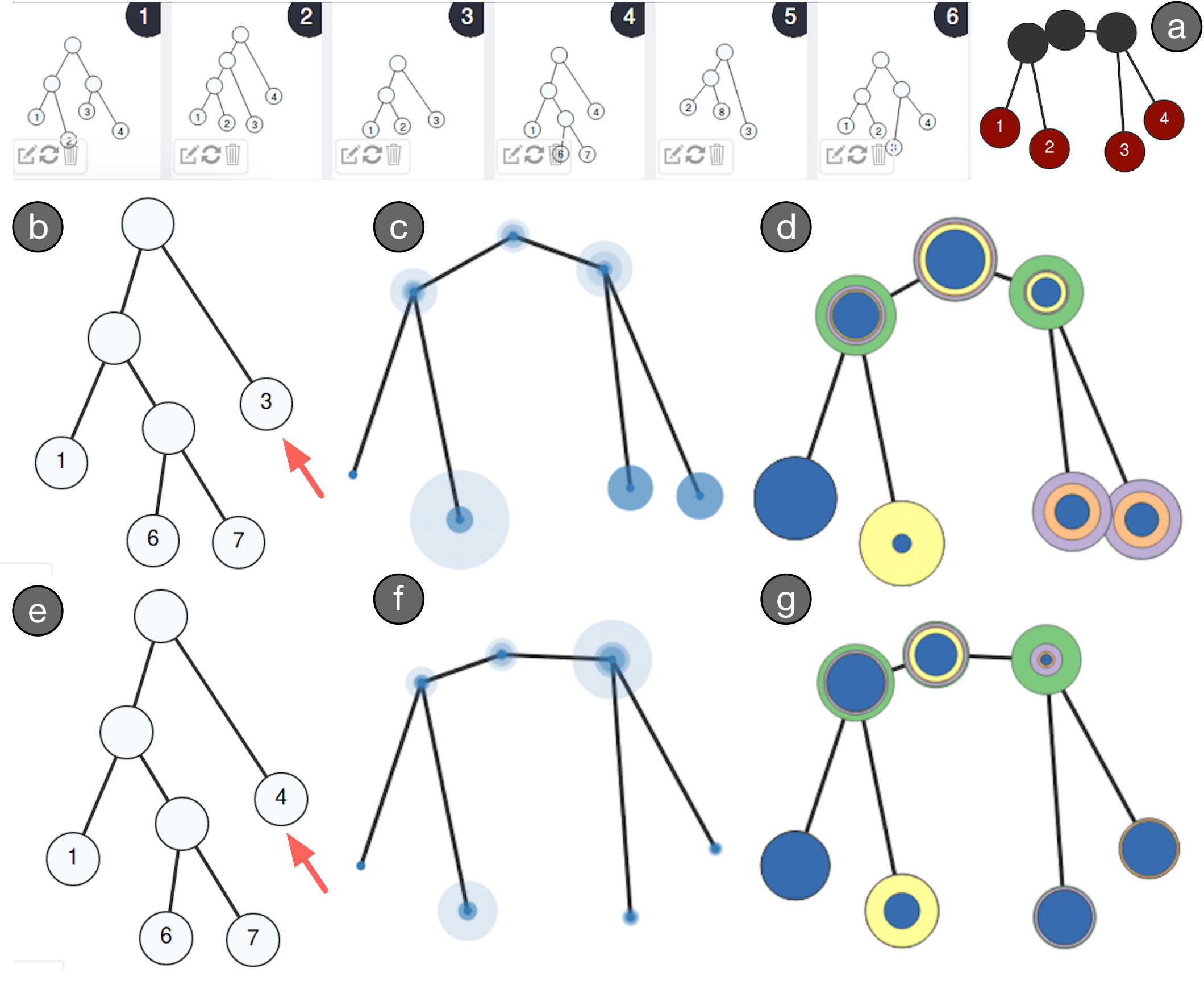}
\vspace{-3mm}
\caption{Label diagnostics and correction. Given an input ensemble of six partially labeled merge trees, by changing a possibly inaccurate initial label from 3 (b) to 4 (e) in \emph{Tree 4}, we reduce the structural variation for the 1-center, comparing (c) and (f). }
\vspace{-3mm}
\label{fig:label-diagnostics}
\end{figure}

We give an example in Fig.~\ref{fig:label-diagnostics}.
The initial input provided by the user contains an ensemble of six leaf-labeled merge trees in partial agreement.
Upon close inspection of the variational consistency plot, we notice that leaves $3$ and $4$ in the 1-center have high variations in their consistency (c).
However, changing a (possibly inaccurate) label in \emph{Tree 4} from $3$ to $4$ greatly reduces the variational consistency of the 1-center at leaves $2$, $3$, and $4$, comparing (c) with (f).
Meanwhile, such a correction also improves the statistical consistency, as shown in (g) where summary statistics (minimum, medium, etc.) coincide at these leaves; see the supplementary video for a demo.
However, such a manual intervention is not practical for large merge trees. We believe the same strategy could be applied to a simplified merge tree on a coarser level. For larger trees, we leave it to future work to automatically identify outliers and suggest rules for modification.

\subsection{Animation Along A Geodesic Path}
Furthermore, we could understand structural similarities in a dynamic setting via animations.
An animated sequence between an ensemble member and the 1-center shows how one deforms to the other via a geodesic (by Theorem \ref{theorem:labelintrinsic}), as well as the evolution of consistency measures at individual nodes through this process.
\begin{figure}[b]
\centering
\vspace{-5mm}
\includegraphics[width=0.98\columnwidth]{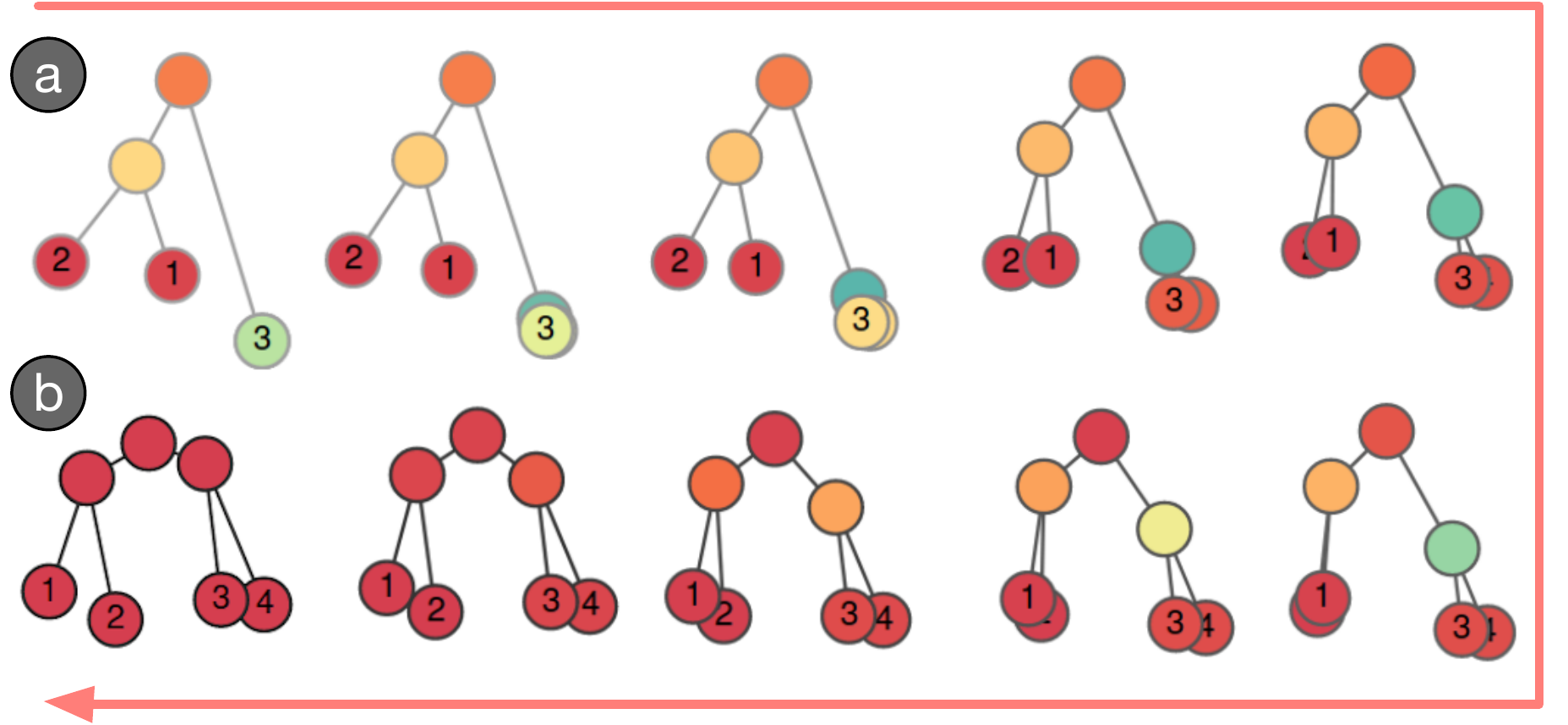}
\vspace{-3mm}
\caption{An animated sequence between an ensemble member (a) and a 1-center tree (b) via a geodesic.}
\label{fig:animation-1}
\end{figure}
 As shown in Fig.~\ref{fig:animation-1}, a 10-step animation between an ensemble member (a) and a 1-center (b) with different leaf sizes  showcases the structural changes between them; e.g., the creation of new internal vertices along the geodesic.
We can also observe the increase in vertex consistencies (red means high and  green means low consistency), in particular, for vertex $3$, along the same geodesic.  See the supplementary video for a demo.

%
%

\section{Application to Scalar Field Ensembles}
\label{subsec:scalarfields}

\begin{figure}[t!]
\centering
\includegraphics[width=0.98\columnwidth]{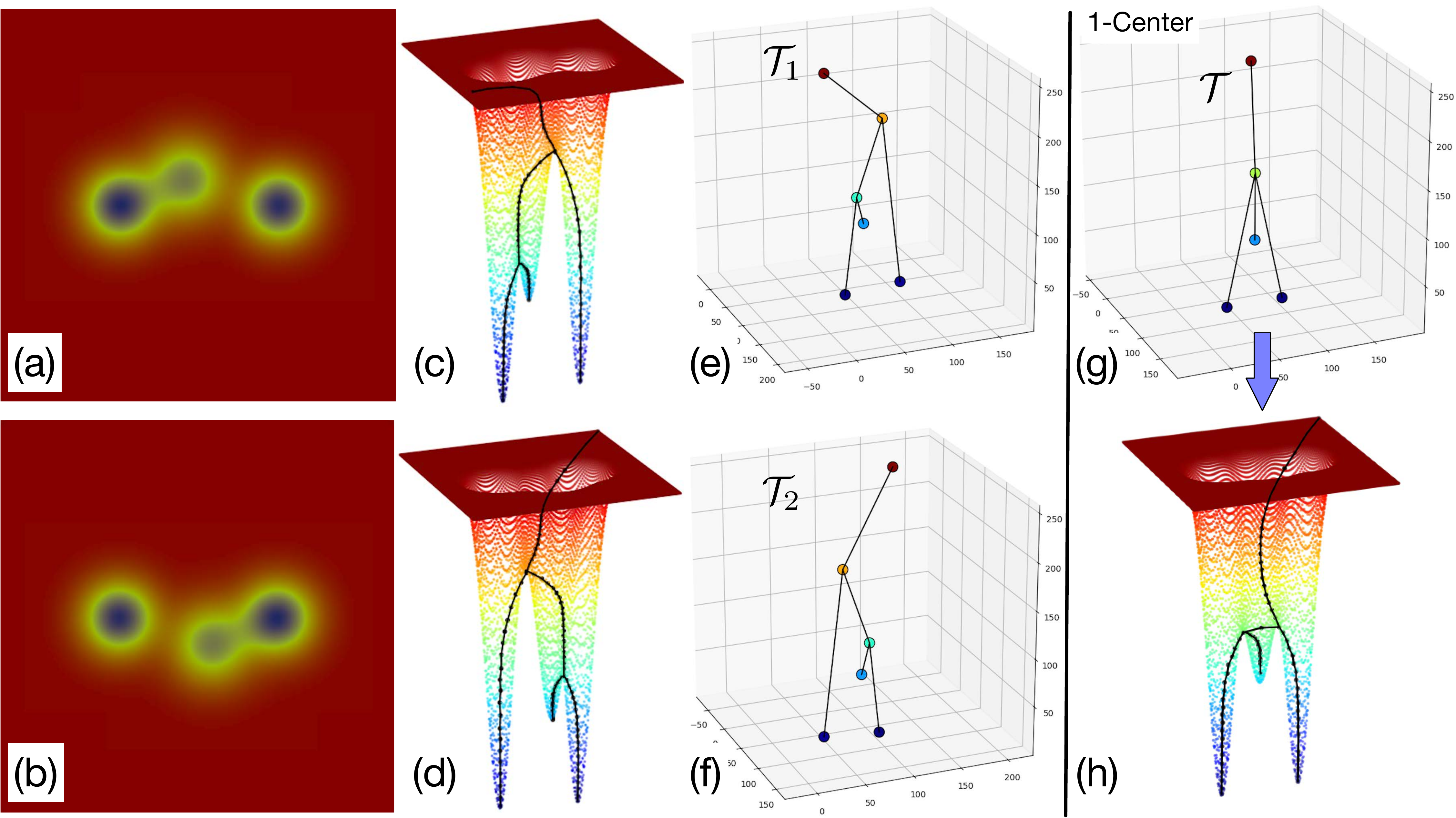}
\vspace{-2mm}
\caption{Computing a 1-center for scalar field induced merge trees. 
On the left, each row contains from left to right: a scalar field visualized using a rainbow color map; an augmented merge tree visualized together with the graph of its corresponding scalar field; and a straight line drawing of a merge tree in $\Rspace^3$. 
On the right: (g) a straight line drawing of the 1-center tree in $\Rspace^3$; (h)  a reverse-engineered scalar field that gives rise to the 1-center tree.}
\vspace{-3mm}
\label{fig:topo-mt-example-1}
\end{figure}

Our framework can be applied to study the structural variation of an ensemble of  scalar fields. We give two illustrative examples.
The first example, Fig.~\ref{fig:topo-mt-example-1}, contains a pair of scalar fields that gives rise to a pair of merge trees very similar to Fig.~\ref{fig:lmt-full} in Section~\ref{subsec:full-agree}.
Each scalar field is visualized as a $150 \times 150$ image in (a)-(b).
We use the \emph{Topology ToolKit}~\cite{TiernyFavelierLevine2018} to generate the augmented merge trees in (c)-(d).
Noncritical points are ignored for the 1-center computation as we need only leaves and their lowest common ancestors to compute the ultra matrices.
We employ the labeling strategy described in Section~\ref{subsec:disagree} to find correspondences between the leaves, by taking into consideration their geometric embeddings.

\begin{figure}[b!]
\centering
\vspace{-5mm}
\includegraphics[width=0.98\columnwidth]{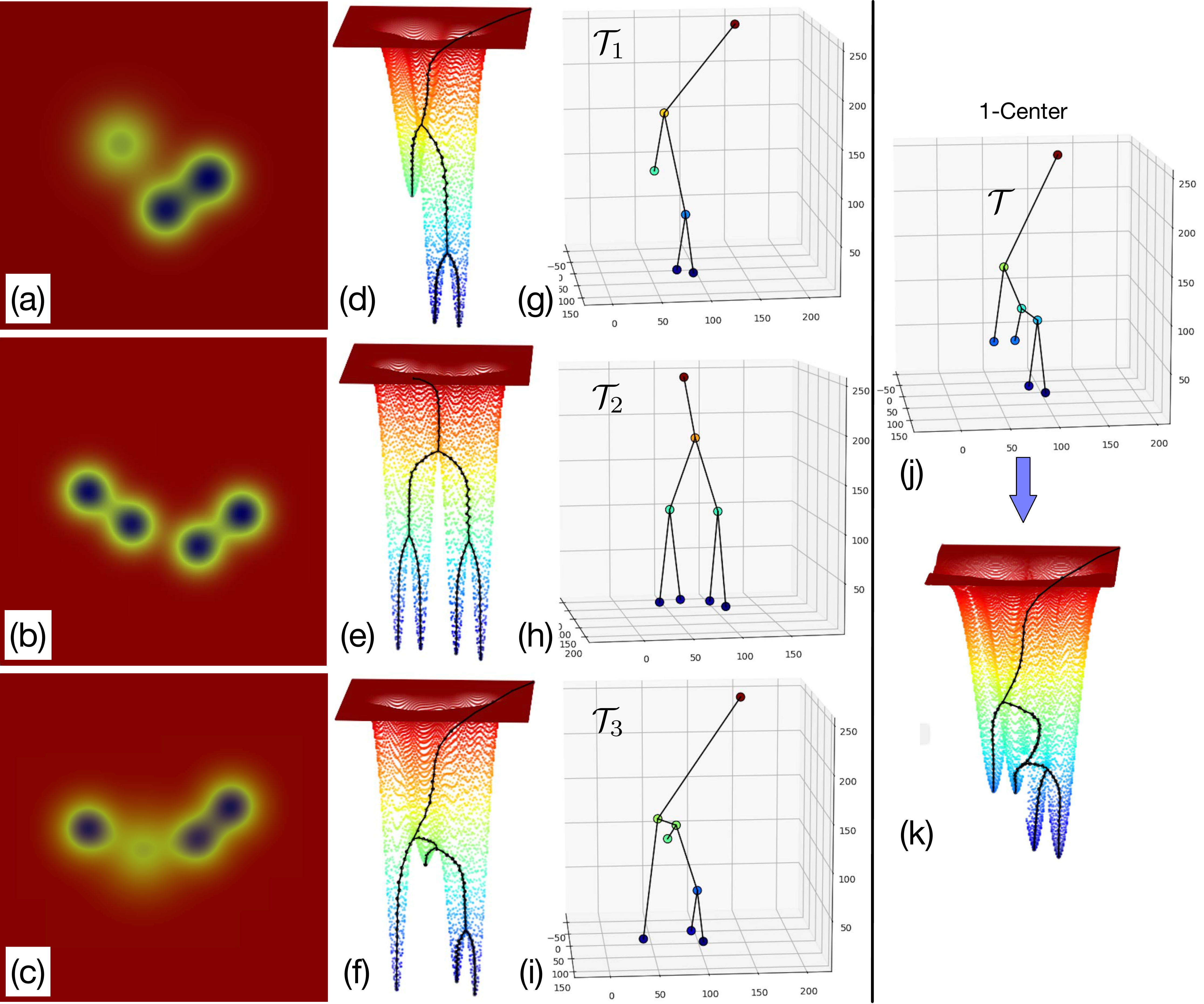}
\caption{Computing a 1-center for merge trees that arise from scalar fields generated by mixtures of Gaussians. Left: each row contains a scalar field visualized as an image, its augmented merge tree,  and a straight line drawing of the corresponding (unaugmented) merge tree in $\Rspace^3$. Right: (j) 1-center tree; and (k) a reverse-engineered scalar field that gives rise to the 1-center tree.}
\label{fig:topo-mt-example-2}
\end{figure}

The second example in Fig.~\ref{fig:topo-mt-example-2} begins with an ensemble of three merge trees that arise from scalar fields generated by mixtures of Gaussians.
We again apply our algorithm in Section~\ref{subsec:disagree} to compute the 1-center tree.
In this example, merge trees $\T_1$, $\T_2$, and $\T_3$ contain $3$, $4$, and $4$ leaves, respectively.
According to Sec.~\ref{subsec:disagree}, $\T_2$ (or $\T_3$) is selected as the pivot tree. We use the Euclidean distance between the geometric embeddings of leaves to assign labels. 
It is easy to verify that the computed 1-center tree $\T$ is indeed of equal distance to the ensemble members $\T_1$, $\T_2$, and $\T_3$, and therefore represents a structural average of the scalar field ensemble.

\begin{figure}[t!]
\centering
\includegraphics[width=0.98\columnwidth]{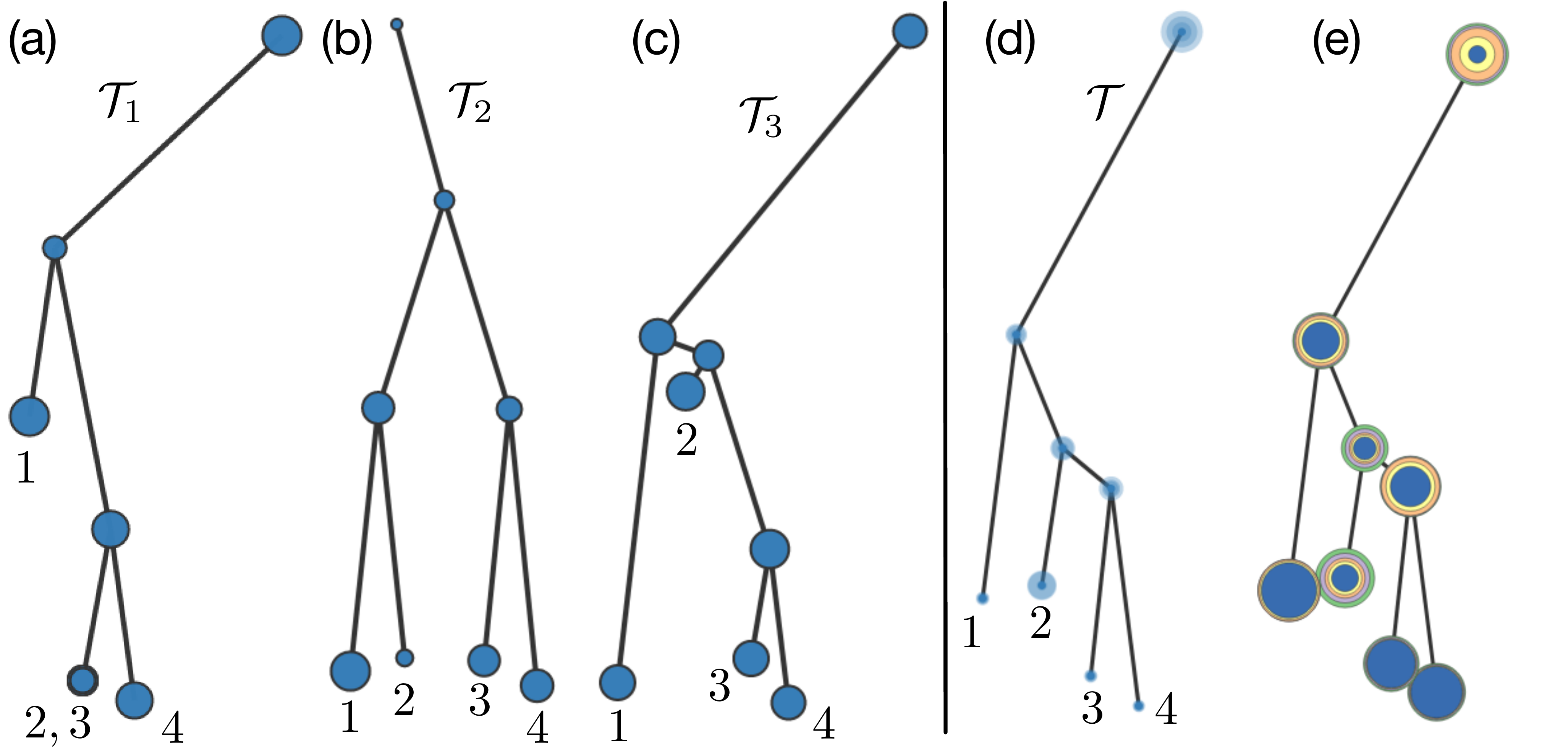}
\vspace{-3mm}
\caption{(a)-(c): Vertex consistencies for each input tree. Leaf labels are inferred based on Euclidean distances among geometric embeddings of vertices. (d) Variational and (e) statistical vertex consistency plot for the 1-center tree.}
\label{fig:topo-mt-example-2-uncertainty}
\vspace{-3mm}
\end{figure}

We now focus on uncertainty visualization using our encodings proposed  in Sec.~\ref{sec:uncertainty} and interactive visualization system in Sec.~\ref{sec:use-cases}.
As illustrated in Fig.~\ref{fig:topo-mt-example-2-uncertainty}, the vertex consistency plot (b) for $\T_2$ indicates that leaf $2$ has low consistency against the 1-center tree; similarly in the variational vertex consistency plot (d), leaf 2 in $\T$ has the highest variation whereas leaves 1, 3, and 4 have low variations across ensemble members.   
This can be explained as all three input trees share similar local structures surrounding leaves 1, 3, and 4.

In addition to our proposed uncertainty visualization, we could further investigate the scalar field that gives rise to a 1-center tree. 
Such a scalar field is certainly not unique, and we have reverse-engineered a candidate by carefully positioning the critical vertices on the 2D domain with radius basis functions, as shown in Fig.~\ref{fig:topo-mt-example-1}(h) and Fig.~\ref{fig:topo-mt-example-2}(k), respectively.   
It is part of our on-going work to reverse-engineer \emph{good} scalar fields  that give rise to a given 1-center tree using optimization techniques with various constraints.

\section{Conclusion and Discussion}
\label{sec:discussion}

We provide an interactive visualization system that computes and visualizes a structural average of an ensemble of leaf-labeled merge trees.
We develop a novel measure of uncertainty, referred to as \emph{consistency},  via a metric space view of the input trees.
This measure is flexible in allowing a local-global tradeoff in understanding structure variations.
Our results are the first steps toward statistical analysis of as well as uncertainty visualization for an ensemble of complex topological descriptors such as merge trees.

There are many future directions.
From an algorithmic perspective, we are interested in computing a 1-center that is robust to outliers, or an effective algorithm to compute 1-mean/1-median. We also want to have a systematic investigation of various heuristic labeling algorithms. 
We are investigating ways to extend our framework to compute 1-center contour trees; as a contour tree of a function $f$ can be constructed by carefully combining merge trees of $f$ and $-f$.
From a visualization perspective, we will evaluate the effectiveness of various visual encodings for consistency measures, as well as the visualization of 1-center trees at scale. 
We will also explore other applications of the animation between merge trees, e.g., in studying shape morphologies in computer graphs.

Finally, we hope that our visualization framework could help enhance topology- and geometry-based modern tools (e.g.,~\cite{LiWangAscoli2017,KanariDlotkoScolamiero2018}) in neuron morphology analysis for biologists; see Appx.~\ref{appendix:results} for an example.
We are also working on helping simulation scientists to take advantage of our framework in detecting abnormalities and outliers within  simulation ensembles.

\acknowledgments{
YW was partially supported by NSF CCF-1740761, RI-1815697, DMS-1547357, and NIH R01EB022899.
EM was partially supported by NSF CCF-1907591 and CMMI-1562012. 
LY and BW were partially supported by NSF DBI-1661375, IIS-1513616 and NIH 1R01EB022876. We thank participants of Dagstuhl Seminar 17292: Topology, Computation and Data Analysis in July 2017 for stimulating discussions on Reeb graphs. We are grateful to the Institute for Computational and Experimental Research in Mathematics (ICERM) for supporting us through the Collaborate@ICERM program in August 2018.
Youjia Zhou created the reverse-engineered scalar fields in Figs.~12 and 13. 
}

\newpage
\bibliographystyle{abbrv-doi}
\bibliography{vis-mean-mt}
\newpage
\pagebreak

\appendix 
\section{From Leaf-Labeled to Labeled Merge Trees}
\label{appendix:internal}

\begin{figure}[h!]
\centering
\includegraphics[width=0.8\columnwidth]{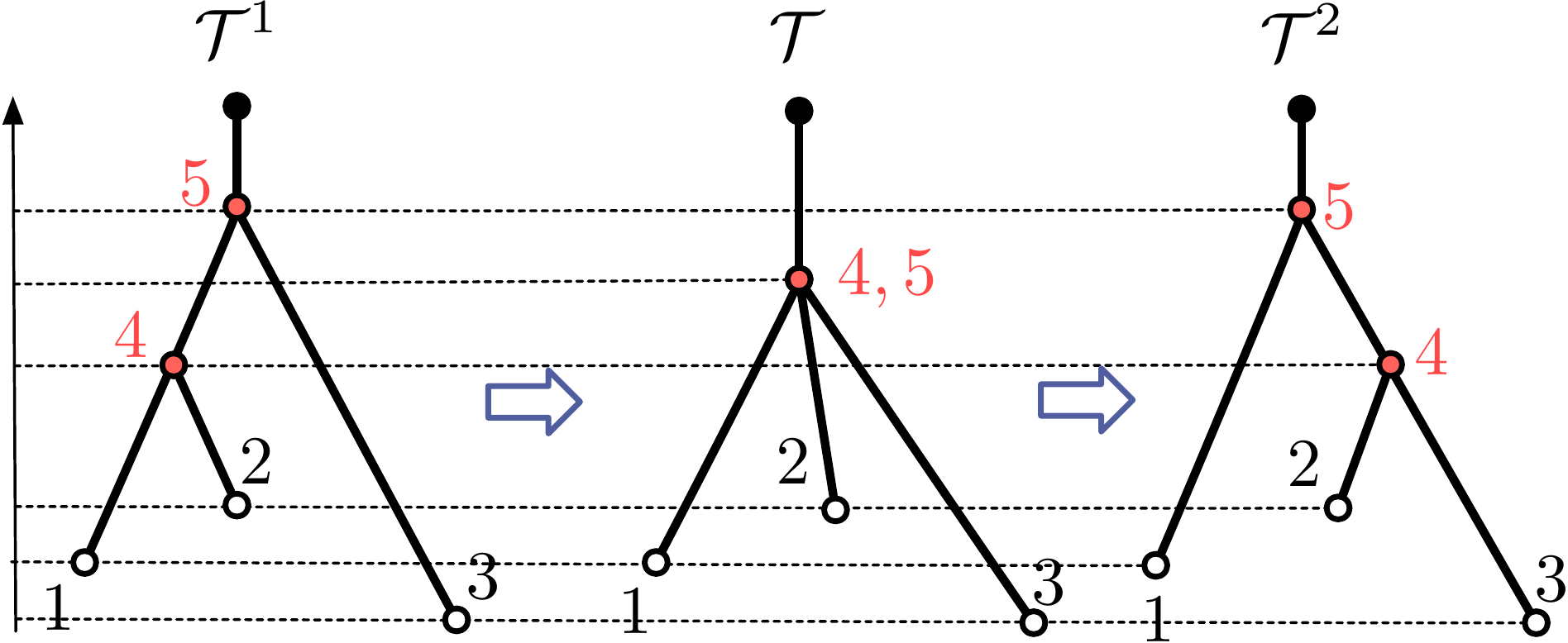}
\vspace{-2mm}
\caption{Inferring a complete labeling for internal vertices based on leaf labels and minimum cost matching.}
\label{fig:lmt-internal}
\end{figure}

As consequences of the algorithms in Section~\ref{sec:implementation}, we arrive at an ensemble of (updated) leaf-labeled merge trees together with their 1-center, which are in full agreement with respect to a shared leaf label set $S$.
For visual embeddings and animations (Sec.~\ref{sec:use-cases}), we need to further infer a 1-to-1 correspondence between internal vertices between $\T^i$ and $\T$ for each $i$. That is, we infer a complete labeling for internal vertices.
Our algorithm is as follows:
\begin{itemize}\denselist
\item [S1.] Transform $\T$ into a pivot tree $\T^p$. Let $S_p$ be its label set.
\item [S2.] For each element $\T^i$, update its labeling using $S_p$.
\end{itemize}
We give a toy example in Figure~\ref{fig:lmt-internal}.

In Step S1, we select a tree $\T^j$ with the largest number of vertices. If $\T^j = \T$, then $\T^p:=\T$. Otherwise, add dummy vertices to $T$ with respect to $T^j$ so that it becomes a tree $T'$ with the largest number of vertices, therefore transforming $\T = (T, f, \omega)$ into a pivot tree $\T^p = (T_p, f_p, \omega_p)$, where $\omega_p: S_p \to V$ is an extension of $\omega: S \to L$ that is surjective on the vertices.

In Step S2, we run the algorithm for trees in partial agreement (Sec.~\ref{subsec:partial-agree}) so that each $\T^i$ is updated to be $\T^{i'} = (T_i, f_i, \omega'_i)$, where  $\omega'_i: S_p \to V_i$ is surjective on its vertices.

\section{Uncertainty Visualization: Edge Consistency}
\label{appendix:edge-consistency}

\para{Edge consistency.}
The vertex consistency can be extended to edge consistency.
For a vertex consistency function $\alpha: V \to \Rspace$ defined on the vertex set $V$ of $\T$, it can be extended to be an edge consistency function using piecewise-linear (PL) $\beta^{PL}: |E| \to \Rspace$ or piecewise-constant (PC) $\beta^{PC}: |E| \to \Rspace$ interpolations in the usual way.
For instance, the PC interpolation of $\alpha$ on an edge takes the minimum value of the two vertex consistencies.

\begin{figure}[!h]
\centering
\includegraphics[width=0.8\columnwidth]{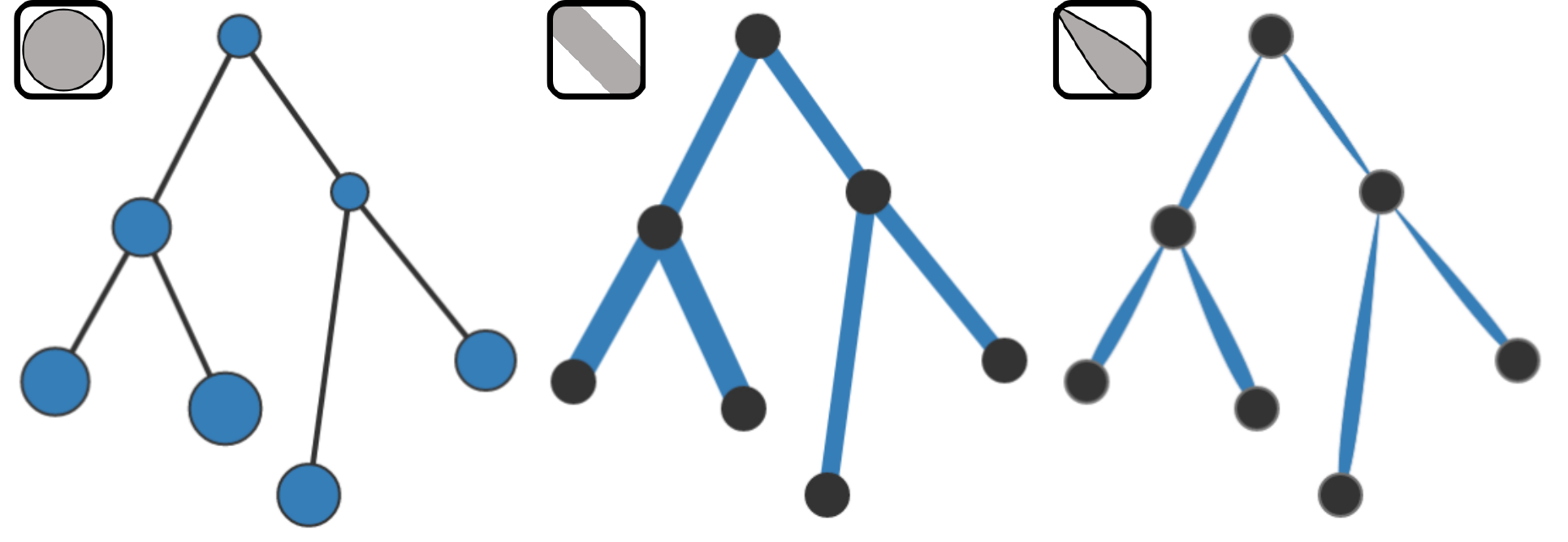}
\vspace{-4mm}
\caption{From left to right: circular, line and ribbon glyphs are used to encode vertex and edge consistencies for an ensemble member.}
\label{fig:glyphs}
\end{figure}

\para{Vertex and edge consistencies for an ensemble member.}
We encode edge consistency for an ensemble member $\T^i$ (with respect to the 1-center $\T$) using glyphs, as in Fig.~\ref{fig:glyphs}.

The width of each line (resp. ribbon) glyph at a location $x \in e$ for $e \in E_i$ scales proportionally with the PC (resp. PL) edge consistency at $x$,
$\beta^{PC}(x)$ (resp. $\beta^{PL}(x)$).

\begin{figure}[!h]
\centering
\includegraphics[width=0.8\columnwidth]{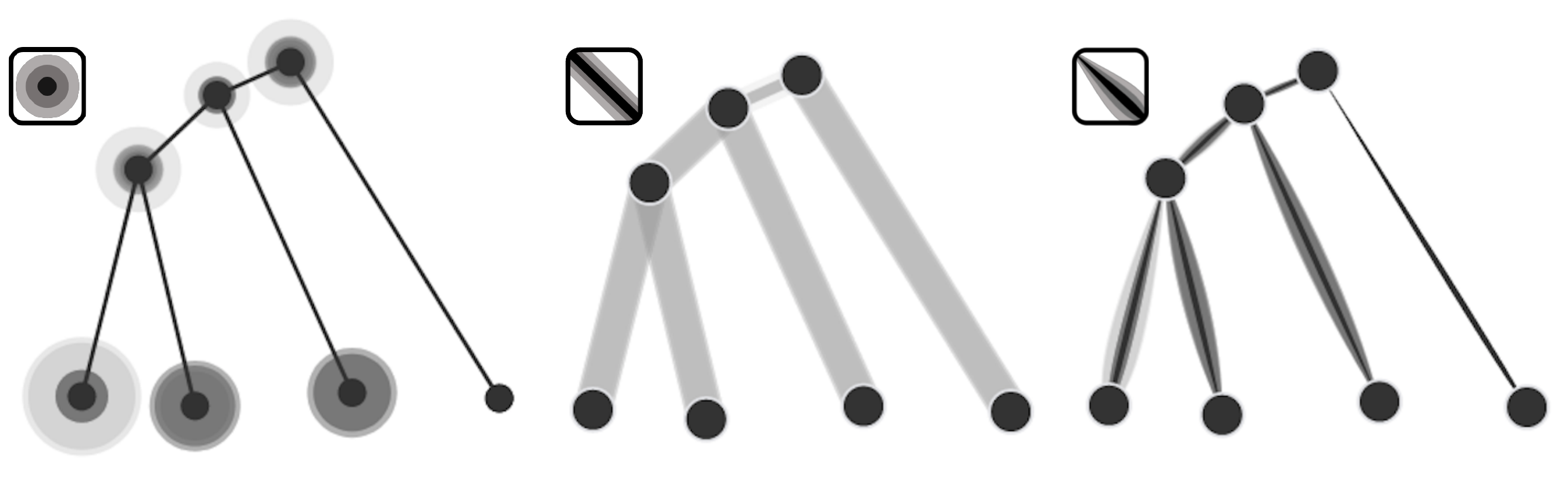}
\vspace{-4mm}
\caption{From left to right: graduated circular glyphs,  graduated lines and  graduated ribbons are used to encode variational vertex, PC and PL edge  consistencies for the 1-center tree.}
\label{fig:glyphs-summary}
\end{figure}

\para{Variational consistencies for the 1-center tree.}
We encode variations in edge consistencies for the 1-center tree $\T$ using visual primitives inspired by~\cite{SanyalZhangBhattacharya2009,SanyalZhangDyer2010}, as in Fig.~\ref{fig:glyphs-summary}, similarly to variations in vertex consistencies.

\begin{figure}[!h]
\centering
\includegraphics[width=0.85\columnwidth]{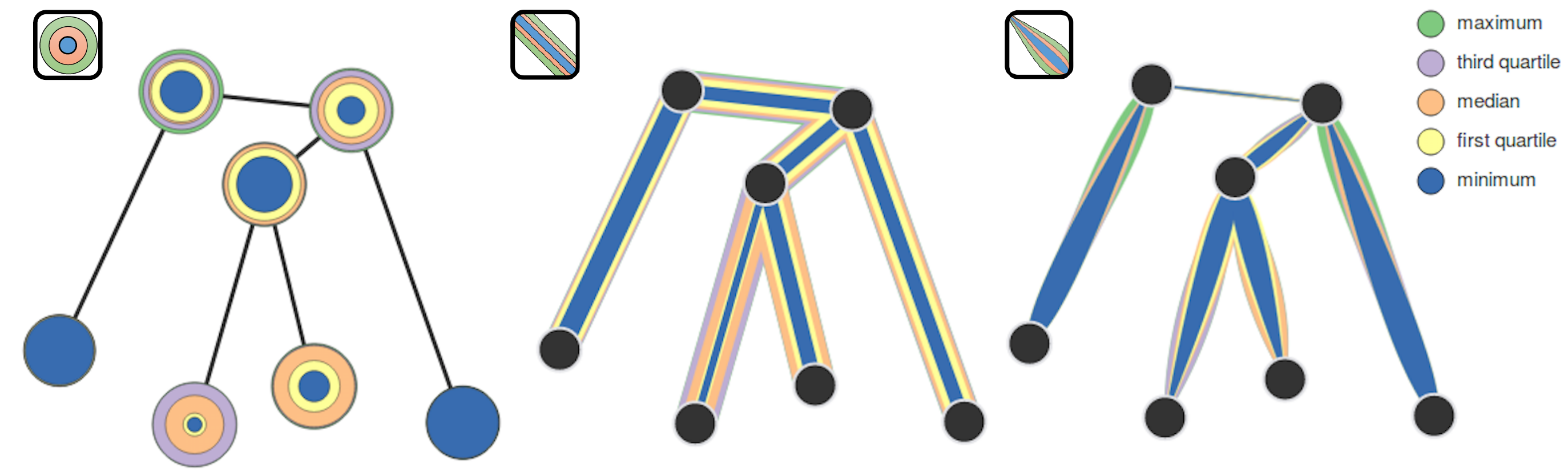}
\vspace{-4mm}
\caption{From left to right: graduated circular glyphs,  graduated lines and  graduated ribbons are used to encode statistical vertex , PC and PL edge consistencies for the 1-center.}
\label{fig:glyphs-statistics}
\end{figure}

\para{Statistical consistency for the 1-center tree.}
Inspired by box plots, we visualize the distribution of edge consistencies at the 1-center $\T$ similarly to the statistical vertex consistencies, see Fig.~\ref{fig:glyphs-statistics}. 

\section{Interactive Visualization System Design Details}
\label{appendix:design}

We provide design details of our interactive visualization system. Its user interface is shown in Fig.~\ref{fig:interface}.

\para{Drawing Panel and Ensemble Panel.}
The drawing panel allows the user to draw individual merge trees using node-link diagrams and assign initial labels to the vertices.
Each tree is created with an embedding onto the drawing panel, where each vertex $v$ is equipped with a coordinate $(v_x, v_y)$ and a height function value $f(v)$ according to the panel's underlying grid structure.
Using various hot keys (see \img{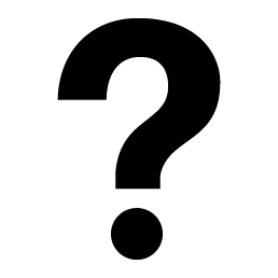} for a user manual), an embedding can be geometrically reconfigured via  insertion, deletion and movement of vertices and edges.
At the moment, the system focuses on leaf-labeled merge trees, therefore only the labels on the leaves are used in the computation.
Each tree is then added \img{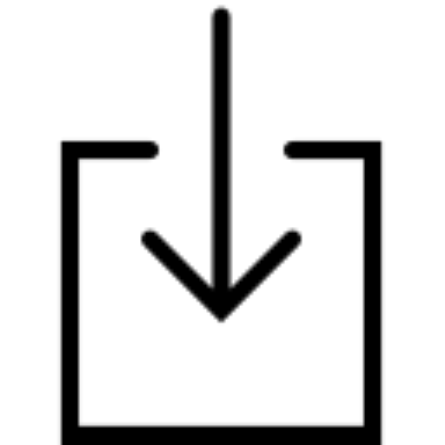} to the ensemble panel, where ensemble members can be selected, deleted and reconfigured/edited \img{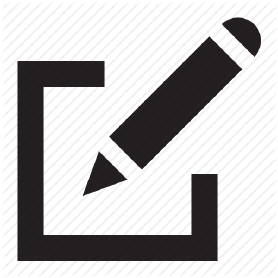}.

\para{Control Panel and 1-center computation.}
With an input ensemble of leaf-labeled merge trees, we compute its 1-center using various options above the control panel.
Using the \myemph{Enforce label} option, we compute a 1-center of trees in full or partial agreement, whereas the \myemph{Ignore label} option enables one to deal with trees in disagreement.

In terms of the parameter setting, we can choose between \myemph{Tree distance} $d_T$ and \myemph{Euclidean distance} $d_E$, or a linear combination of the two, using the $\bm{\lambda}$ parameter for our heuristic labeling strategies.
$\bm{\delta}$ is the locality parameter in consistency measures and {$\# \textbf{steps}$} indicates the number of steps used in the animation.

The control panel also visualizes the relation between the 1-center (denoted as a red node labeled \myemph{AMT}) and the ensemble members in a star-shaped summary plot, where the 1-center lies in the center of the star, and the color and length of each link scales proportionally with the interleaving distance between an input tree and the 1-center.
By clicking on a link in the summary plot, we enable an animated sequence between an input tree and the 1-center.

\para{Animation.}
We compute and visualize an animated sequence between an input tree (as the source) and the 1-center (as the target) using two strategies. The \myemph{geodesic strategy} \img{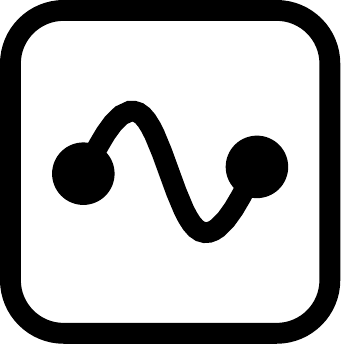} follows Theorem~\ref{theorem:labelintrinsic}, where intermediate trees follow a geodesic connecting the source and the target pair.
The \myemph{linear strategy} \img{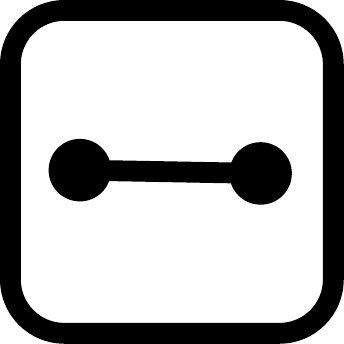} linearly interpolates between geometric embeddings of the source and the target.

During the animation, the intermediate trees can be displayed with \myemph{labels} \img{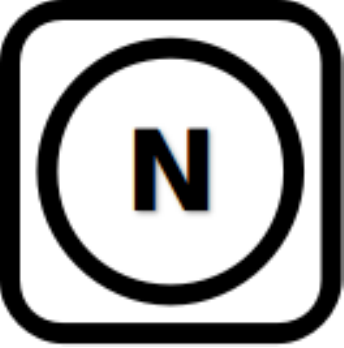}, which highlight correspondences between leaves.
The intermediate trees can also use \myemph{colored labels} \img{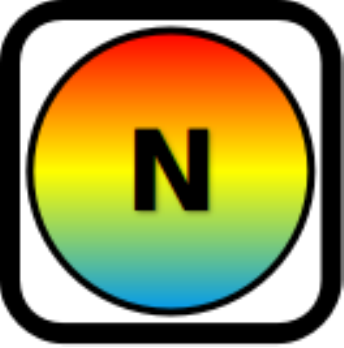} according to vertex consistencies, for which the animated  sequence highlights the changes in vertex consistency as a source tree is moved towards a target tree.
In summary, the animation highlights structural variations between an input tree  and the 1-center, and the evolution in vertex consistency during such a process.

\para{Consistency Visualization.}
We visualize various consistency measures in the rightmost panel.
First, we visualize vertex and edge consistency for each ensemble member with respect to the 1-center, using circular \img{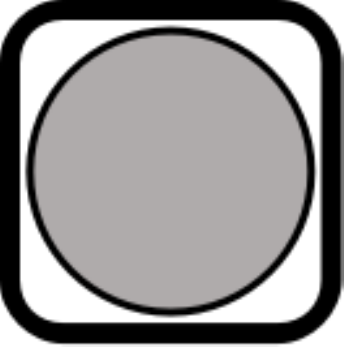}, linear \img{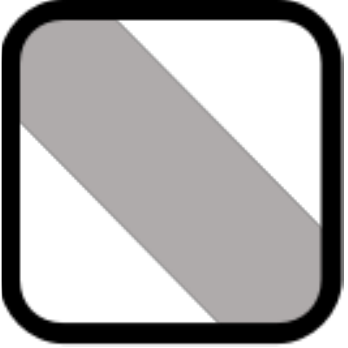} and ribbon glyphs \img{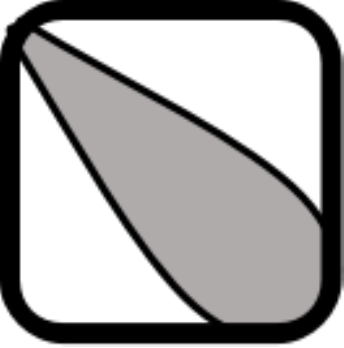}.
Second, we visualize variational consistencies for the 1-center using graduated  glyphs with sequential colormap of a single hue.
We use graduated circular glyphs \img{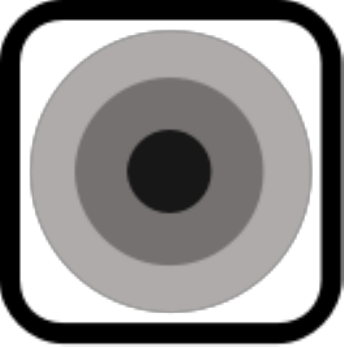} for vertices, graduated lines \img{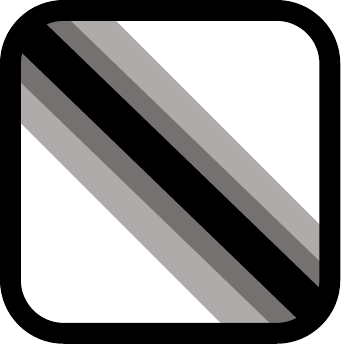} and graduated ribbons  \img{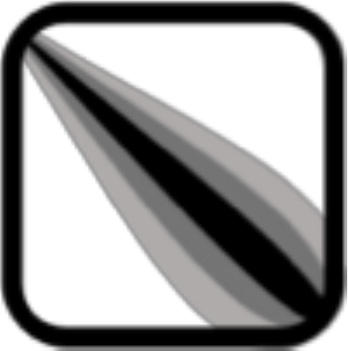} for PC and PL edges, respectively.
Finally, we highlight statistical consistencies for vertices \img{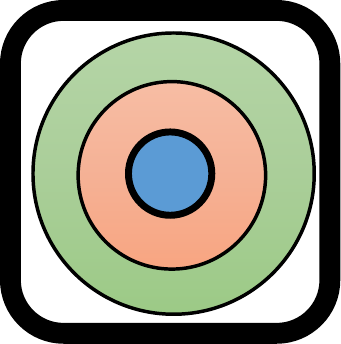}, PC edges \img{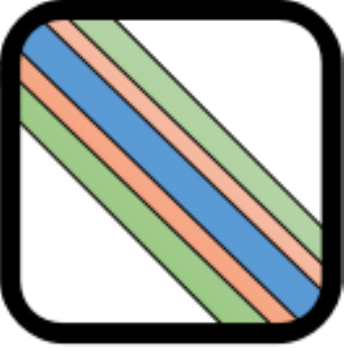} , and PL edges \img{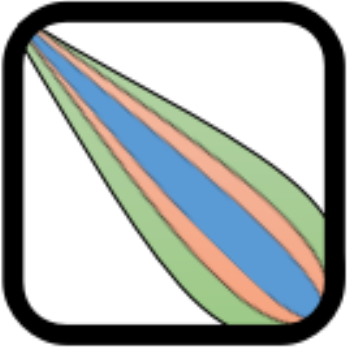} that capture the distribution of consistency measures across ensemble members.
In addition, all trees can be visualized with \myemph{labels} \img{symbol-label.pdf} to indicate leaf correspondences between input and output.

\para{1-Center tree visualization.}
Since each input tree $\T^i$ is drawing with a geometric embedding $\iota_i: |T_i| \to \Rspace^2$, we compute an embedding $\iota$ of the 1-center $\T$ using information from $\iota_i$ for all $1 \leq i \leq k$.
First we apply the algorithm in Sec.~\ref{sec:implementation} and Appx.~\ref{appendix:internal} to infer a complete correspondence between the internal vertices of $\T^i$ and $\T$ for each $i$.
Then, for each vertex $v \in V$ with a label $l$, we compute its embedding $\iota(v) \mapsto (x_v, y_v)$. $y_v$ comes naturally as $y_v = f(v)$. $x_v$ is the 1-center of the x-coordinates of $\{\iota_1(\omega_1(l)), \cdots, \iota_k(\omega_k(l))\}$.

\section{Local-Global Tradeoff}
\label{appendix:localglobal}

\begin{figure}[h!]
\centering
\includegraphics[width=0.98\columnwidth]{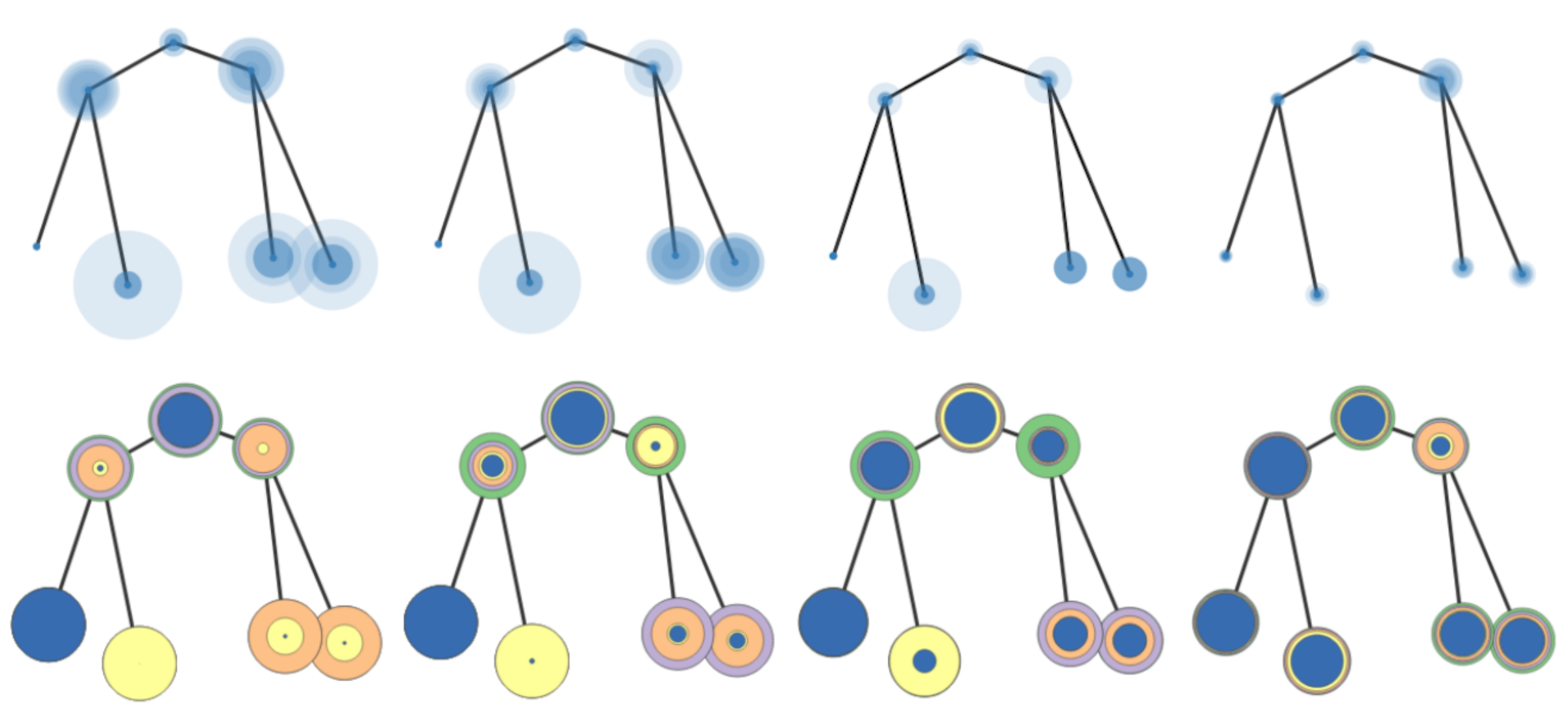}
\vspace{-2mm}
\caption{The changes in variation and distribution of consistency as we increase the smoothing parameter $\delta \in \{0.05,0.07,0.10,0.15\}$.}
\label{fig:change-delta}
\end{figure}

We could also use our system to study the tradeoff between local and global consistency measures in capturing structural similarities between ensemble members and their structural average.
As we increase the locality parameter $\delta$ in the Gaussian-weighted cosine similarity measure, we could observe the change in structural variations on vertices; see Fig.~\ref{fig:change-delta} for an example.
Using the same input ensemble as Fig.~\ref{fig:label-diagnostics}, we see that as $\delta$ increases, the variational vertex consistency decreases (top) as we pay more attention to global structural similarities among the ensemble members.
Meanwhile, the distribution of consistency measures becomes increasingly concentrated (bottom). See the supplementary video for a demo.

\section{Intrinsic-Extrinsic Tradeoff}
\label{appendix:inextrinsic}

The input merge trees can have natural (function-induced) intrinsic metrics associated to them. Sometimes, these trees are geometric (i.e., embedded in Euclidean spaces), and thus also have natural extrinsic (ambient) metrics defined on them, e.g., in the case of neuron trees modeling neuron cells.
Our tool supports a combination of both metrics. See the supplementary video for a demo.

Fig.~\ref{fig:usage-intrinsic} shows that using purely Euclidean distance vs. purely tree distance gives different new labels for \emph{Tree 2} and  \emph{Tree 3} and thus affects the resulting 1-center as well as its statistical consistency.
The results using a linear combination of both Euclidean and tree distance for $\lambda = 0.5$ are very similar to those based on pure tree distance, with minor differences visible for the 1-center trees and their statistical consistency plots.

\begin{figure}[h!]
\centering
\includegraphics[width=0.8\columnwidth]{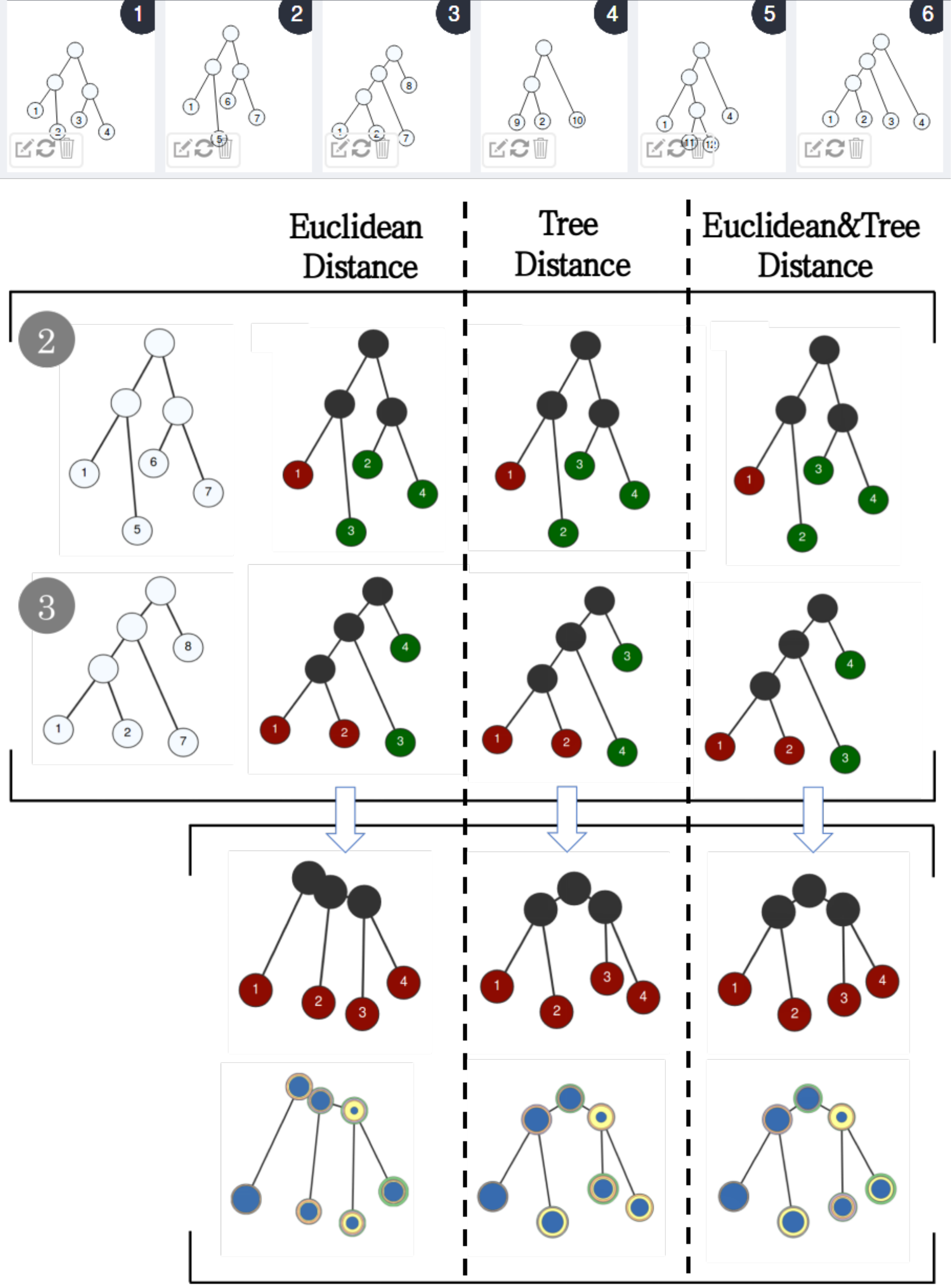}
\vspace{-2mm}
\caption{Investigating the tradeoff between intrinsic and extrinsic metrics in labeling strategies. 1st row: input ensemble. 
2nd row, from left to right: \emph{Tree 2}; \emph{Tree 2} with updated labels using Euclidean distance, tree distance, and a combination of both using  $\lambda=0.5$. 
3rd row, similar to 2nd row, for  \emph{Tree 3}. 
4th and 5th row: 1-center tree and its statistical consistency plot; from left to right: using Euclidean distance, tree distance, and a combination of both using $\lambda=0.5$.}
\label{fig:usage-intrinsic}
\end{figure}

\section{Applications in Neuron Morphology}
\label{appendix:results}
Neuron cells have tree morphology, and a rapidly increasing amount of neuroanatomical data are now publicly available (e.g., \url{NeuroMorpho.org} and \url{flycircuit.org}). Our proposed consistency measure can be used to understand structural variations among an ensemble of \emph{neuron cell induced merge trees} with respect to their 1-center.

As a case study, we use our proposed methodologies to help study differences/variations among different reconstructions of the same neuron cell.
In particular, in the past 15 years, a large number of algorithms have been developed to reconstruct a tree structure for neuron cells (referred to as \emph{neuron trees}) from 2D or 3D images (e.g.,~the dozens of methods incorporated in the visualization software Vaa3D \cite{vaa3d}).

\para{Input trees.}
We use one of the olfactory projection fibers datasets, referred to as $\OP$, from the DIADEM challenge~\cite{BrownBarrionuevoCanty2011}.
We create a set of neuron trees reconstructed for $\OP$ using different reconstruction methods.
Each neuron tree comes with a 3D embedding in the form of a 3D image.
Given an image of a neuron tree, we extract its corresponding (unlabeled) merge tree representation as a pair $\T^i  = (T_i, f_i)$. First, the vertex set $V_i$ is obtained by extracting the leaves and branching points from the 3D image; each vertex is equipped with a geometric  location via the 3D embedding of the neuron tree. 
The function $f_i: V_i \to \Rspace$ is the geodesic distance of a vertex $x \in V_i$ to a base point $o \in V_i$ (chosen as the physical root of the neuron cell). This extracted merge tree is referred to as the \emph{neuron cell induced merge tree}.

\begin{figure}[t!]
\centering
\includegraphics[width=.99\columnwidth]{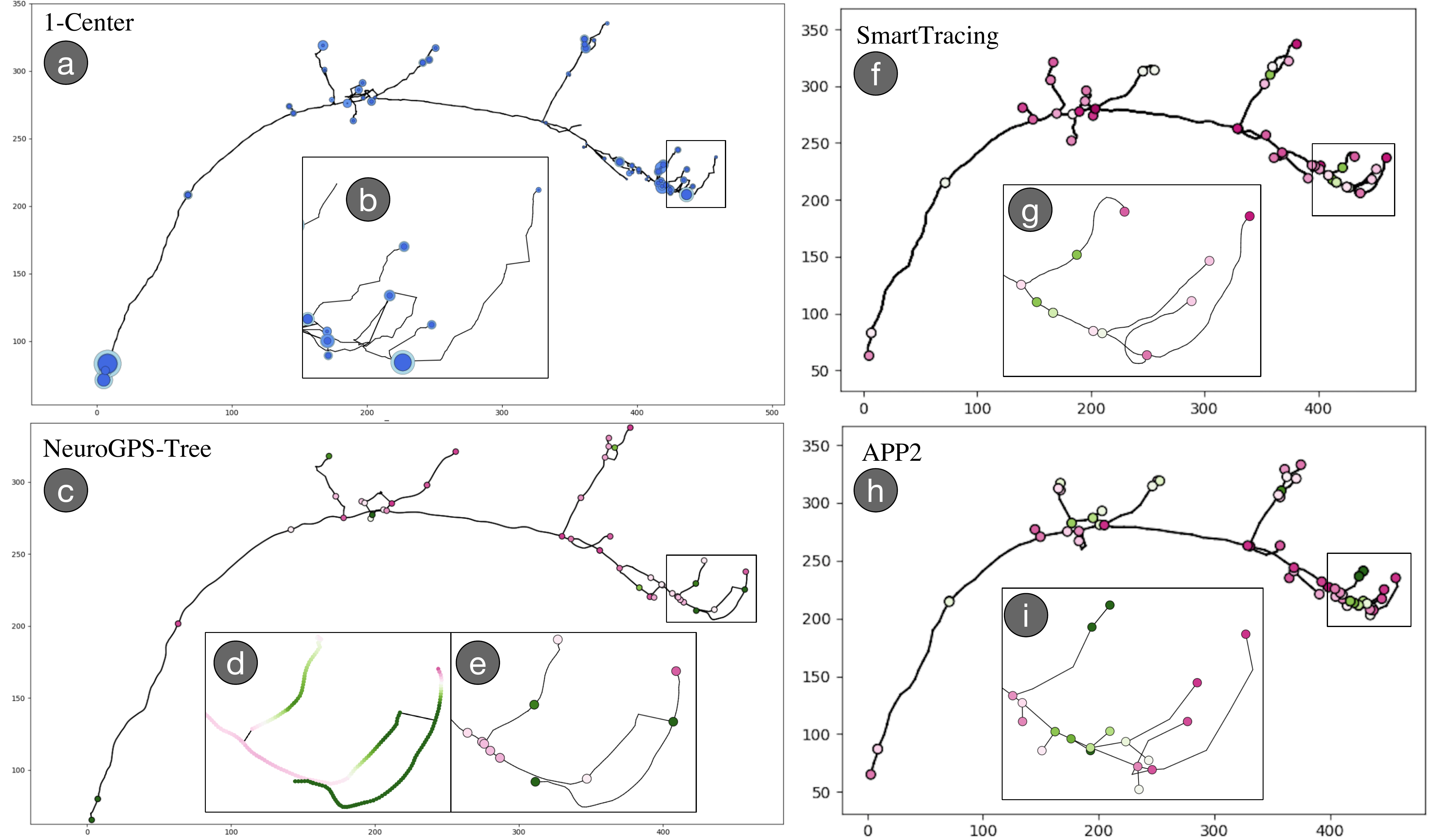}
\vspace{-2mm}
\caption{2D projections of vertex consistency visualization of an ensemble of  neuron trees reconstructed from $\OP$.}
\vspace{-2mm}
\label{fig:op6}
\end{figure}

\para{Analysis.}
Fig.~\ref{fig:op6} shows three neuron cell induced merge trees for dataset $\OP$ via reconstruction algorithms APP2~\cite{XiaoPeng2013} (h), SmartTracing~\cite{ChenXiaoLiu2015} (f) and NeuroGPS-Tree~\cite{QuanZhouLi2016} (c), as well as their 1-center tree (a).
The vertices for each input tree are colored by their consistencies.
The input trees are reasonably similar; thus, to see the structural variation, we use a relatively small $\delta = 0.05$ value when computing consistency.

Meanwhile, each vertex of the 1-center tree is visualized with a variational consistency using graduated circular glyphs; see Fig.~\ref{fig:op6} (c,f,h).
It is therefore easy to spot which vertices in the 1-center tree have high variance.
Upon close inspection, each high variance vertex in the 1-center tree indeed  corresponds to locally different reconstructions in input trees (e,g,i).
Similarly, from each individual input tree, it is easy to see how each vertex  deviates from the 1-center tree locally. For example, vertices from the region in the black rectangle of NeuroGPS-Tree have low consistency (white to green colors) with respect to the 1-center tree (e).
Indeed, as the inset zoomed-in view shows (e), NeuroGPS-Tree produces a different local tree configuration (one with a severely bended branch) as the output of APP2 (i) and SmartTracing (g).
We also remark that as we increase the $\delta$ value, the consistency measures similarity at a more global level and thus the structural variation becomes less visible.
Finally, we can also show consistency along edges of input trees, which could help to make low-consistency regions more prominent to spot than the vertex consistency visualization. See Fig.~\ref{fig:op6}(d) for an example.
A detailed methodological development for neuron trees  is left for future work.

\end{document}